\documentclass[a4paper,11pt]{article}
\pdfoutput=1
\usepackage{jcappub}

\usepackage{amsmath,mathtools}
\usepackage{stmaryrd}
\usepackage{tabularx}
\usepackage{soul}
\usepackage{booktabs}
\usepackage{multirow}
\usepackage{comment}
            
\usepackage{xspace}

\newcommand{\lcdm}{$\Lambda$CDM\xspace}

\newcommand{\apjl}{Astrophys. J. Lett.}

\newcommand{\apjs}{Astrophys. J. Suppl. Ser.}
\newcommand{\aap}{Astron. Astrophys.}

\defcitealias{musso_2012_correlated}{MS}
\defcitealias{sheth_mo_tormen_2001}{SMT}
\defcitealias{SVdW}{SVdW}
\defcitealias{jennings2013}{Vdn}

\title{The universal multiplicity function: counting haloes and voids}

\author[a,b,c,d]{Giovanni Verza,} 
\author[e]{Carmelita Carbone.} 
\author[b,f,g]{Alice Pisani,}
\author[h]{Cristiano Porciani,}
\author[c,d,i,j]{and Sabino Matarrese}

\affiliation[a]{Center for Cosmology and Particle Physics, Department of Physics, New York University, 726 Broadway, New York, NY 10003, USA}
\affiliation[b]{Center for Computational Astrophysics, Flatiron Institute, 162 5$^{\rm th}$ Avenue, 10010, New York, NY, USA}
\affiliation[c]{Dipartimento di Fisica e Astronomia ``G. Galilei",
Universit\`a degli Studi di Padova, via Marzolo 8, I-35131 Padova, Italy}
\affiliation[d]{INFN, Sezione di Padova, via Marzolo 8, I-35131 Padova, Italy}

\affiliation[e]{INAF – Istituto di Astrofisica Spaziale e Fisica cosmica di Milano (IASF-MI), Via Alfonso Corti 12, I-20133 Milano, Italy}

\affiliation[f]{The Cooper Union for the Advancement of Science and Art, 41 Cooper Square, New York, NY 10003, USA}
\affiliation[g]{Department of Astrophysical Sciences, Peyton Hall, Princeton University, Princeton, NJ 08544, USA}

\affiliation[h]{Argelander-Institut für Astronomie, Auf dem Hügel 71, D-53121 Bonn, Germany}

\affiliation[i]{INAF - Osservatorio Astronomico di Padova,
Vicolo dell’Osservatorio 5, I-35122 Padova, Italy}
\affiliation[j]{Gran Sasso Science Institute, Viale F. Crispi 7, I-67100 L’Aquila, Italy
}

\abstract{
\noindent
We present a novel combination of the excursion-set approach with the peak theory formalism in Lagrangian space and provide accurate predictions for halo and void statistics over a wide range of scales. The set-up is based on an effective moving barrier. Besides deriving the corresponding numerical multiplicity function, we introduce a new analytical formula reaching the percent level agreement with the exact numerical solution obtained via Monte Carlo realisations down to small scales, $\sim 10^{12} h^{-1}{\rm M_\odot}$.
In the void case, we derive the dependence of the effective moving barrier on the void formation threshold, $\delta_{\rm v}$, by comparison against the Lagrangian void size function measured in the DEMNUni simulations. We discuss the mapping from Lagrangian to Eulerian space for both haloes and voids; adopting the spherical symmetry approximation, we obtain a strong agreement at intermediate and large scales.
Finally, using the effective moving barrier, we derive Lagrangian void density profiles accurately matching measurements from cosmological simulations, a major achievement towards using void profiles for precision cosmology with the next generation of galaxy surveys. 
}

\begin{document}

\maketitle

\section{Introduction}
Our knowledge of the underlying cosmological model of the Universe 
relies on the analysis of summary statistics of observed data. 
One standard way to theoretically model these statistics is to compute the distribution of over- and under-densities in the initial matter density field, the so-called Lagrangian space, linearly evolved down to the epoch of interest\footnote{Note that in the excursion-set and peak theory frameworks the initial density is intended as the matter density field at a redshift high enough to consider it still in the linear regime.}. In this perspective, the two main approaches are the excursion-set formalism~\citep{bond_et_al1991} and the theory of Lagrangian density peaks~\citep{bardeen_1986}, which have been widely explored to provide theoretical models for the dark matter (DM) halo mass function (HMF), bias, halo correlation functions~\citep[][and references therein]{press_schechter_1974,bardeen_1986,Peacock1990,bond_et_al1991,Lacey_Cole1993,mo_white_1996,porciani_matarrese1998,catelan_1998,sheth_tormen_1999,maggiore_riotto_I_2010, musso_2012_correlated}, and cosmic void size functions (VSF)~\citep{SVdW}, hereafter \citetalias{SVdW},~\citep{jennings2013,achitouv_2015}.

On the one hand, to model in the initial density field the distribution of regions that would form haloes or voids in the evolved Universe, the excursion-set approach considers the density contrast field filtered, i.e. smoothed, on various scales at random positions, which can be described by a random walk~\citep{Peacock1990,bond_et_al1991}. haloes and voids are considered formed when the random walks reach the formation threshold, also called barrier. The typical quantity that can be obtained is the multiplicity function, which is the first-crossing distribution of the forming barrier at a given scale. From this quantity, the HMF, the VSF, and the corresponding local bias expansion can be derived~\citep[see][]{desjacques_bias}. 

On the other hand, the formalism associated to the peak theory models the progenitors of haloes (and voids) using local maxima (minima) in the initial density field, filtered at some scale~\citep{bardeen_1986}. 
However, in this case the link between the number density of peaks and the corresponding multiplicity function is not straightforward. A way to obtain a multiplicity function from peak theory is to combine the statistics of Lagrangian density peaks with the excursion-set~\citep{paranjape_2012_pk-es}.

In this work we explore an effective way to combine the excursion-set framework with peak theory, for both halo and void statistics. In particular, we show that this can be obtained using the standard excursion-set framework with an effective formation barrier, which is a moving barrier that does not correspond to the physical formation threshold, but it can capture the statistical properties of haloes and voids. We show that this approach can be used to model the multiplicity function and even higher order statistics, such as the density profile.
Our approach is valid for both haloes and voids; however, here we mainly focus on voids statistics, as in this case a consistent theoretical modelling is still missing.

This paper is organised as follows. In Sec.~\ref{sec:ex-set_general} we review the building blocks of the excursion-set formalism and excursion-set of peaks, revisiting the halo and void formation within these frameworks. In particular, we propose an effective way to merge the excursion-set with peak theory via an effective moving barrier; in Sec.~\ref{sec:multipl_func} we discuss the exact numerical solution of our new theoretical model and derive an accurate analytical approximation; 
in Sec.~\ref{sec:effective_barrier} we discuss how to obtain the effective barrier and derive the one for cosmic voids; in Sec.~\ref{sec:eulerian_mapping} we introduce the Lagrangian to Eulerian mapping for both haloes and voids; in Sec.~\ref{sec:profile} we apply our model to compute the Lagrangian density profile of voids; finally we draw our conclusions in Sec.~\ref{sec:conclu}.

\section{haloes and voids in Lagrangian space}\label{sec:ex-set_general}

The excursion-set approach is based on the concept of modelling the statistical properties of haloes or voids in Lagrangian space and mapping them to the fully nonlinearly evolved density field, the Eulerian space. 

The fundamental quantity of the excursion-set formalism is the smoothed, linearly evolved initial density field at a comoving coordinate $\mathbf{q}$, 
\begin{equation}\label{eq:filtered_delta}
\begin{split}
\delta(\mathbf{q},R) &= \int \mathrm{d}^3 x \, W(|\mathbf{x}|,R) \delta(\mathbf{q}+\mathbf{x}) \\
&= \int \frac{\mathrm{d}^3 k}{(2 \pi)^3} W(k R) \delta(\mathbf{k}) e^{-i\mathbf{k \cdot q}}\,.
\end{split}
\end{equation}

Halo and void formation in the excursion-set are described by a thresholding process of the filtered Lagrangian field. A halo with a Lagrangian radius $R$ (and corresponding mass $M(R)$) is formed at the Lagrangian position ${\bf q}$ if $R$ is the maximum smoothing scale at which the filtered density matter field, $\delta({\bf q},R)$, crosses the halo formation threshold~\citep{Peacock1990,bond_et_al1991}. 
Void formation is analogous to the halo case, with the additional condition that the filtered field that crosses the (negative) formation threshold has not crossed the (positive) halo formation threshold at any larger scale~\citepalias{SVdW}. 
This shows that the treatment of halo and void formation in the excursion-set framework requires the following quantities: formation threshold, filtering function, and position.

\paragraph{\textbf{Halo and void formation thresholds}}A halo is a gravitationally bound, virialized object; for this reason, the threshold for halo formation is chosen as the linear density contrast corresponding to the full collapse in the Eulerian space~\citep{bond_et_al1991}. 
At a first approximation, the dynamics of the halo collapse can be incorporated through a scale-dependent threshold, usually called ``moving barrier''~\citep[\citealp{sheth_mo_tormen_2001},~hereafter \citetalias{sheth_mo_tormen_2001};][]{,sheth_tormen_2002,robertson_2009,paranjape_lam_shet_2012,elia_2012,sheth_2013,ludlow_2014,borzyszkowski_2014}.

Contrary to the halo case, cosmic voids do not experience any particular event in their evolution, which may characterize their formation: an underdense region starts to grow faster with respect to the background and continues its outward expansion forever (in the single-stream regime). 
It follows that a map from the Lagrangian to Eulerian density contrast of filtered underdense fluctuations always exists. 
This condition ensures that the properties (density and size) of evolved voids can always be mapped in the linear theory, and {\it vice versa}. 
In this perspective, the linear void formation threshold $\delta_{\rm v}$ can be any negative value~\citep{pisani_2015_abundance, nadathur_2015,verza_2019,euclid_vsf}, that can be mapped in the corresponding Eulerian one.

\paragraph{\textbf{The filter function}}The filter function $W(k R)$, entering Eq.~\eqref{eq:filtered_delta}, is statistically related to the halo and void definitions in Lagrangian space~\citep{bond_et_al1991}.  
In addition, the detailed statistical properties of $\delta({\bf q},R)$ depend on the choice of the specific filter function; the most commonly used ones are the top-hat, Gaussian, and sharp-$k$ filters~\citep[see][for other physically motivated filter functions]{bardeen_1986}. In this work we consider the top-hat filter, which in Fourier space is defined as $W(kR) = 3 j_1(kR)/kR$, where $j_1(x)=(\sin x - x \cos x)/x^2$ is the spherical Bessel function of order 1. 
This filter has a clear physical interpretation and a well-defined associated Lagrangian mass, $M_{\rm TH} =  4 \pi R^3 \rho_{\rm m} / 3$, where $\rho_{\rm m}$ is the mean comoving matter density~\citep[see][for details on other filters]{porciani_matarrese1998}.

\paragraph{\textbf{Lagrangian field positions}} In the classical formulation of the excursion-set model, the Lagrangian position, ${\bf q}$, of the density contrast field is random, and the corresponding derived statistics for haloes and voids are obtained weighting over the entire Lagrangian space. 
It follows that the corresponding multiplicity function describes the fraction of the Lagrangian volume where the filtered field $\delta({\bf q},R)$ (first) reaches the threshold value. Mapping this quantity to the number density of discrete objects is not straightforward. 
However, massive haloes and voids form and evolve around special positions, such as, respectively, maxima and minima in Lagrangian space~\citep{robertson_2009,ludlow_porciani_2011,massara2018}. 
In the literature, it was shown that solving the first crossing distribution problem over all the Lagrangian positions leads to a mismatch between the moving barrier measured in cosmological simulations from the distribution of the Lagrangian patches that will form haloes in the evolved universe, and the one obtained by fitting the excursion-set multiplicity function against the measured HMF~\citep[\citetalias{sheth_mo_tormen_2001},][]{robertson_2009}. 
To fix this issue, Paranjape and Sheth~\citep{paranjape_2012_pk-es} proposed an approximated method to derive the multiplicity function from the excursion-set over the subset of maxima rather than over the entire space. They found that at large scales (where the adopted approximations are accurate enough), they can recover the difference between the barriers obtained in the two different ways.

The focus of this work is to further explore the excursion-set on the subset of Lagrangian extremants, in order to provide a theoretical description of the \textit{halo and void} statistics.
To to this aim, we introduce the following  definition: a Lagrangian halo (void) with radius $R$ and the filtered density contrast $\delta({\bf q},R)$ satisfying the following conditions: \\ 
i) the Lagrangian position ${\bf q}={\bf q}(R)$ is not contained in any larger halo (void);
ii) the Lagrangian position ${\bf q}={\bf q}(R)$ is a maximum (minimum) of the Lagrangian density field filtered at the smoothing length $R$; 
iii) $R$ is the largest scale at which $\delta({\bf q},R)$ crosses the positive (negative) formation threshold, without having crossed, in the case of voids, the positive halo formation threshold at any larger scale. Note that condition i) generalizes the cloud-in-cloud exclusion of the excursion-set.

In principle, it is possible to derive the exact solution of various statistics by combining the excursion-set with the peak theory. However, this is complicated both analytically and computationally, and only approximate solutions are practically feasible~\citep{paranjape_2012_pk-es,paranjape_2013}. On the other hand, the comparison between the approximated excursion-peak multiplicity function of Paranjape and Sheth~\cite{paranjape_2012_pk-es} with the (approximated) excursion-set one of Musso and Sheth~\citep{musso_2012_correlated} (hereafter \citetalias{musso_2012_correlated}) and \citetalias{sheth_mo_tormen_2001}, implicitly suggests an alternative approach to face this problem: the excursion-peak multiplicity function can be derived from the standard excursion-set one by using an effective moving barrier, which however does not correspond to the physical density contrast of Lagrangian haloes and voids~\citep{achitouv_2013}. 
In this work, we show that such an effective barrier contains the statistical information of the multiplicity function and beyond.
As a final remark, we underline that in this paper we focus on the void and halo statistics in Lagrangian space; we introduce the mapping from Lagrangian to Eulerian in Sec.~\ref{sec:eulerian_mapping}, nevertheless we leave the full implementation for future work.

\section{The multiplicity function from a generic moving barrier}
\label{sec:multipl_func}

The stochastic evolution of the Lagrangian density contrast field is formally described by the Langevin equations~\citep{Peacock1990, bond_et_al1991, 
porciani_matarrese1998}. 
This field is assumed here to be a Gaussian random field; nevertheless, it is possible to extend the following description to account also for non-Gaussianity~\citep{matarrese_verde_jimenez_2000,matarrese_verde_2008,carbone_2008_PNG,maggiore_riotto_III_2010,achitouv_2012_nonGaussian,musso_2014_nonGaussian}. The effect of varying the smoothing scale $R$ can be obtained by differentiating Eq.~\eqref{eq:filtered_delta}:
\begin{equation}\label{eq:langevin_eq_1}
\frac{\partial \delta(R)}{\partial R} = \int \frac{\mathrm{d}^3 k}{(2 \pi)^3} \frac{\partial W(k R)}{\partial R} \delta(\mathbf{k}) e^{-i\mathbf{k \cdot q}} = \mathcal{Q}(R)\,,
\end{equation}
where $\delta(R)=\delta(\mathbf{q},R)$.
This has the form of the Langevin equation, which shows how an infinitesimal change of the scale $R$ affects the smoothed field as a function of the stochastic force $\mathcal{Q}(R)$. The initial condition of this first-order stochastic differential equation is given by $\delta(R) \rightarrow 0$ as $R \rightarrow \infty$.
Since Eq.~\eqref{eq:langevin_eq_1} is linear, the stochastic force is also a Gaussian random field with a vanishing expectation value, $\langle \mathcal{Q}(R) \rangle=0$, and it is uniquely described by its correlation function. Therefore, the Langevin equation is fully described by the equation system~\citep{porciani_matarrese1998}
\begin{equation}\label{eq:langevin_eq_2}
\begin{cases}
\dfrac{\partial \delta(R)}{\partial R} = \displaystyle\int \dfrac{\mathrm{d}^3 k}{(2 \pi)^3} \dfrac{\partial W(k R)}{\partial R} \delta(\mathbf{k}) e^{-i\mathbf{k \cdot q}} = \mathcal{Q}(R)\\[2ex]
\langle \mathcal{Q}(R_1) \mathcal{Q}(R_2) \rangle = \displaystyle\int_0^\infty \dfrac{\mathrm{d}k \, k^2}{2 \pi^2} P(k) \dfrac{\partial W(k R_1)}{\partial R_1} \dfrac{\partial W^*(k R_2)}{\partial R_2}\,,
\end{cases}
\end{equation}
where $P(k)$ is the linear matter power spectrum. The coherence of each trajectory along $R$ depends only on the form of the filter function.

\subsection{The numerical multiplicity function}To numerically obtain the statistical quantities we are interested in, such as the multiplicity function, we solve the first crossing problem over a large number of realisations of the stochastic Eq.~\eqref{eq:langevin_eq_2}~\citep{bond_et_al1991,robertson_2009,paranjape_lam_shet_2012,achitouv_2015}.
The direct solution is numerically expensive, to speed up the computation we exploit the Gaussianity of the density contrast field plus the Cholesky method~\citep{nikakhtar_2018}. 
Let us consider the correlation of the field $\delta(R)$ filtered at different smoothing lengths $R$
\begin{equation}\label{eq:C_ij_def}
\langle \delta(R_i) \delta(R_j) \rangle \equiv C_{ij} = \int \frac{{\rm d} k \, k^2}{2 \pi^2} P(k) W(kR_i) W^*(kR_j)\,.
\end{equation}
The matrix ${\bf C}$, with elements $C_{ij}$, is real, symmetric, and positive-definite, so it has a unique decomposition ${\bf C} = {\bf L L }^T$, where ${\bf L}$ is a lower triangular matrix. It is therefore possible to construct a stochastic numerical realisation of $\delta(R)$ as
\begin{equation}\label{eq:delta_sigle_cholesky}
\delta(R_i) = \sum_j L_{ij} G_j \,,
\end{equation}
where $G_j$ is the $j^{\rm th}$ element of a vector of Gaussian variables, following a distribution with zero mean and unit variance. The multiplicity function is computed as the fraction of realisations, $N_\times / N_{\rm tot}$, that first cross the formation barrier, $B(R)$, between $R_{i}$ and $R_{i+1}$
\begin{equation}\label{eq:num_f}
(R_{i}-R_{i+1}) f_{R_{i},R_{i+1}} = N_\times[B(R_{i+1})] / N_{\rm tot}.
\end{equation}

\subsection{A new analytical multiplicity function}The exact analytical solution of the first crossing distribution exists only for a few cases, which are the ones with uncorrelated or fully-correlated steps with a constant or linear evolving threshold~\citep{bond_et_al1991,sheth_1998,SVdW,paranjape_lam_shet_2012,musso_2012_correlated}. The first crossing distribution for the general case can in principle be written as a formal expansion in an infinite series of functions involving $n-$point correlations of the field $\delta$ and its derivatives evaluated at different scales~\citep{verechtchaguina_2006,musso_2014_nonGaussian,nikakhtar_2018}, but summing this series requires approximations. Therefore, approximated methods must be used to analytically solve the Langevin equations in the excursion-set framework with correlated steps and a generic moving barrier~\citep{Peacock1990,apple_1990,riotto_2010,de_simone_2011,ma_maggiore_2011,paranjape_lam_shet_2012,paranjape_2012_pk-es,musso_2012_bias,musso_2014_markVel,musso_2013}.

In the standard cosmological model, the variance of the fluctuation field, $S=\sigma^2(R)=C(R,R)$ (where $C(R,R)$ is a diagonal element of the covariance matrix defined in Eq.~\eqref{eq:C_ij_def}), is a monotonic function of the smoothing scale $R$; therefore, these two quantities are interchangeable.
To account for correlations, \citetalias{musso_2012_correlated} proposed a multiplicity function derived from a bivariate distribution of the density contrast filtered field and of the velocity of the filtered field with respect to the scale $S$, $\delta' = {\rm d} \delta(S) / {\rm d} S$. 
They replace the requirement of $\delta(s) < B(s)$ for all the scales $s<S$ (with $B(s)$ a generic moving barrier) by the milder condition $\delta(S) - \Delta S < B(S - \Delta S)$, with $S-s=\Delta S  \rightarrow dS \rightarrow 0$. 
In this limit, expanding in a Taylor series both the field $\delta(S)$ and the moving barrier $B(S)$, the above condition becomes
\begin{equation}
B_S \leq \delta_S \leq B_S + (\delta_S' - B_S') \Delta S , \quad \delta_S' \geq B_S',
\end{equation}
with $\Delta S \rightarrow 0$, where $\delta_S = \delta(S)$, $\delta_S' = \delta'(S)$, $B_S = B(S)$, and $B_S' = B'(S)$. The random walk distribution is now approximated by the joint distribution $p(\delta_S,\delta'_S)$, and the corresponding multiplicity function reads
\begin{align}\label{eq:f_MS}
\begin{split}
f(S) \, {\rm d} S &\simeq \int_{B'_S}^\infty {\rm d} \delta' \int_{B_S}^{B_S + (\delta_S' - B_S') {\rm d} S } {\rm d} \delta_S\, p(\delta_S,\delta_S') \\
&= {\rm d} S \, p(B_S) \int_{B'_S}^\infty {\rm d} \delta' p(\delta'_S|B_S) (\delta_S' - B_S')\,,
\end{split}
\end{align}
where $ p(\delta_S'|B_S) = p(B_S,\delta_S') / p(B_S) $ is the conditional probability of $\delta_S'$ at the barrier $\delta_S=B_S$~\citepalias{musso_2012_correlated}. The resulting multiplicity function accurately reproduces the exact numerical solution for a general filter at large scales $R$, i.e. at small $S$, nevertheless it becomes less accurate when lowering the scale. 
This is expected since the correlation between different smoothing scales is modelled for an infinitesimal distance from the crossing scale, so as a local term. 
Spanning over a large range of scales, the local term is no longer sufficient for an accurate model, since it is not able to capture the behaviour of the field at scales significantly different from the crossing.

One possibility to extend this model consists in exploring the next leading orders of the formal expansion of~\citep{musso_2014_nonGaussian}. Nevertheless, here we explore an alternative way which relies on accounting for correlations of the density contrast field on different scales. More precisely, we consider the joint distribution $p(\delta_S,\delta'_S,\delta_s)$, where $\delta_s=\delta(s)$, $S$ is the scale at which the first crossing occurs and $s<S$. In this way, by accounting for $\delta_S'$, we can model the correlation on scales close to the crossing barrier, as in the \citetalias{musso_2012_correlated}, but accounting also for $\delta(s)$ we can model the behaviour of the field at very different scales. Using the same arguments of \citetalias{musso_2012_correlated}, we obtain
\vspace{-0.18cm}
\begin{align}\label{eq:def_f_new}
f(S) \, {\rm d} S &\simeq \,\frac{1}{S} \int_0^S {\rm d} s \int_{-\infty}^{B(s)} {\rm d} \delta_s \int_{B'_S}^\infty {\rm d} \delta'_S \int_{B_S}^{B_S + {\rm d} S (\delta'_S - B'_S)} p(\delta_S,\delta'_S,\delta_s ) {\rm d} \delta_S \\\nonumber 
&= \frac{{\rm d} S}{S} \int_0^S \hspace{-0.5em}{\rm d} s \int_{-\infty}^{B(s)} \hspace{-0.5em} {\rm d} \delta_s \int_{B'_S}^\infty \hspace{-0.5em} {\rm d} \delta'_S 
\, p\big(B_S,\delta'_S,\delta_s \big)  (\delta'_S - B'_S),
\end{align}
where in the last line $\delta_S=B_S$.
The integration over $\delta_S$, $\delta'_S$, and $\delta_s$ represents the probability that the field $\delta$ crosses the barrier $B$ at scale $S$, without having crossed it at the larger scale $s$. 
Integrating over $s$, we then obtain the probability that the field $\delta$, crossing the barrier at $S$, has not crossed the barrier at any scale $s<S$. The $1/S$ term is the normalisation factor: $p(\delta_S,\delta'_S,\delta_s)$ is normalised to $1$ at each $s$, therefore its integration over $s$ from 0 to $S$ gives $S$. As the field is Gaussian, the joint distribution of $\delta_S$, $\delta'$, and $\delta_s$ is a multivariate normal distribution with covariance
\vspace{-0.085cm}
\begin{equation}\label{eq:covariance}
{\bf \Sigma} = \begin{pmatrix}
\langle \delta_S^2 \rangle & \langle \delta_S \delta'_S \rangle & \langle \delta_S \delta_s \rangle\\
\langle \delta_S \delta'_S \rangle & \langle \delta'^2_S \rangle & \langle \delta'_S \delta_s \rangle\\
\langle \delta_S \delta_s \rangle& \langle \delta'_S \delta_s \rangle & \langle \delta_s^2 \rangle
\end{pmatrix}= \begin{pmatrix}
S   & 1/2 & C \\
1/2 & D   & C' \\
C   & C'  & s 
\end{pmatrix}.
\end{equation}
Here we used the fact that, by construction $\langle \delta_S^2 \rangle=S$, $\langle \delta_s^2 \rangle=s$, $\langle \delta_S \delta_s \rangle=C(S,s)$ defined in Eq.~\eqref{eq:C_ij_def}, that from now on we call simply $C$. We also define the quantities $D\equiv\langle \delta'^2_S \rangle$ 
and $C'\equiv\langle \delta'_S \delta_s \rangle=\partial C / \partial S$. It can be shown that $\langle \delta_S \delta'_S \rangle = 1/2$ (see Appendix~\ref{ap:properties}).
Differently from previous works, here we evaluate exactly the quantity  $\langle \delta'^2_S\rangle$. 
In fact, for a of a top-hat filter, this quantity is usually considered non-convergent, due to the slowly decreasing amplitude and highly oscillating functions to integrate. In Appendix~\ref{ap:compute_W} we discuss the convergence of $\langle \delta'^2_S \rangle$, showing how to properly compute it. 
Using the properties listed in Appendix~\ref{ap:properties}, we obtain 
\begin{align}\label{eq:Ps_final}
{\cal P}(s) =& \int_{-\infty}^{B(s)} {\rm d} \delta_s \int_{B'}^\infty {\rm d} \delta'_S p(B_S,\delta'_S,\delta_s)  (\delta'_S - B'_S) = \nonumber \\
&\frac{1}{4 \pi \sqrt{\Gamma_{\delta \delta}}} \frac{\det {\bf \Sigma}}{\Gamma_{\delta' \delta'}} \exp \left[ -\frac{D B_S^2 + SB'^2_S - B'_S B_S}{2 \Gamma_{\delta \delta}} \right] \times \\\nonumber
&\qquad\qquad\left\{ {\rm erf} \left[ \sqrt{\frac{\Gamma_{\delta \delta}}{2 \det {\bf \Sigma}}} \left(B(s) + \frac{\Gamma_{B \delta}}{\Gamma_{\delta \delta} }B_S + \frac{\Gamma_{\delta' \delta}}{\Gamma_{\delta\delta} }B'_S \right) \right] + 1\right\}+ \\\nonumber
&\frac{\Gamma_{\delta' \delta}}{4 \pi S \sqrt{\Gamma_{\delta' \delta'}}} \exp \left(-\frac{S B^2(s) + sB_S^2 - 2 CB(s)B_S}{2 \Gamma_{\delta' \delta'}} \right) + \\\nonumber 
& e^{-B_S^2/2S}\Bigg\{ \frac{\det {\bf \Sigma}}{2 \pi \Gamma_{\delta' \delta} \sqrt{\Gamma_{\delta' \delta'}}} {\cal F}(x,\alpha,\beta) 
+\\
&\qquad\qquad\frac{B_S/2S - B'_S}{4\sqrt{2 \pi S} } \left[ {\rm erf} \left( \sqrt{\frac{S}{2 \Gamma_{\delta' \delta'}}} \left(B(s) - \frac{C}{S}B_S \right) \right) +1 \right]\Bigg\}
 , \nonumber
\end{align}
with
\begin{equation}\label{eq:gammas}
\begin{array}{lcl}
\Gamma_{BB} = sD - C'^2 & & \Gamma_{B\delta'} = C C' - s/2 \\
\Gamma_{\delta' \delta'} = sS - C^2 &  & \Gamma_{B\delta} = C'/2 - C D \\
\Gamma_{\delta \delta} = SD - 1/4 &  & \Gamma_{\delta' \delta} =C/2 - S C' \,,
\end{array}
\end{equation}
and
\begin{equation}\label{eq:Fxab}
{\cal F}(x,\alpha,\beta) = \int_{-\infty}^x {\rm d}t \, t \, e^{-\alpha(t+\beta)^2} {\rm erf}(t)\,,
\end{equation} where
\begin{align}\label{eq:x_alpha_beta}
x & = \frac{\Gamma_{\delta' \delta}}{\sqrt{2 
\Gamma_{\delta' \delta'} \det {\bf \Sigma}}} \left( B(s) + \frac{\Gamma_{B \delta'}}{\Gamma_{\delta' \delta}} B_S + \frac{\Gamma_{\delta' \delta'}}{\Gamma_{\delta' \delta}} B'_S \right) \,, \nonumber\\
\alpha & = \frac{\Gamma_{\delta \delta} \Gamma_{\delta' \delta'}}{\Gamma_{\delta' \delta}^2}-1 = \frac{S \det {\bf \Sigma}}{\Gamma_{\delta' \delta}^2} \,, \\
\beta & = \sqrt{\frac{\Gamma_{\delta' \delta'}}{2 \det {\bf \Sigma}}} \left( \frac{B_S}{2 S} - B'_S \right) \,.\nonumber
\end{align}
Finally, the analytical multiplicity function is given by integrating Eq.~\eqref{eq:Ps_final}: 
\begin{equation}\label{eq:f_new_final}
f(S) \, {\rm d} S = \frac{{\rm d} S}{S} \int_0^S {\rm d} s \, {\cal P}(s)\,.
\end{equation}
Its computation is numerically straightforward, as the quantities appearing in Eq.~\eqref{eq:Ps_final} are a combination of the elements of the covariance matrix, ${\bf \Sigma}$, which can be easily computed for any filtering function. 
Studying the behaviour of ${\cal P}(s)$, in Appendix~\ref{ap:limit} we show that $\lim_{s\rightarrow 0} {\cal P}(s) = \lim_{s\rightarrow S} {\cal P}(s)$. It follows that, in the small $S$ limit, Eq.~\eqref{eq:f_new_final} can be further approximated as
\begin{align}\label{eq:multipl_integr_approx}
f(S) \simeq \frac{e^{-B^2_S / 2S}}{\sqrt{2 \pi S}}  &\left\llbracket  \sqrt{\frac{\Gamma_{\delta \delta}}{2 \pi S}} \exp \left[ -\frac{S}{2\Gamma_{\delta \delta}} \left(\frac{B_S}{2S} - B'_S \right)^2\right]\right. + \nonumber \\
& \ \left. \frac{1}{2} \left(\frac{B_S}{2S} - B'_S \right) \left\{{\rm erf}\left[\sqrt{\frac{S}{2\Gamma_{\delta \delta}}} \left(\frac{B_S}{2S} - B'_S \right)\right]+1\right\} \right\rrbracket \,. 
\end{align}
This equation is formally the \citetalias{musso_2012_correlated} multiplycity function, with the difference that in this work the $\Gamma_{\delta \delta}$ term, which depends on $D$, is exactly computed accounting for its scale dependence rather than using a fixed approximated value. This approximation is quite accurate at medium and large scales, involving only four quantities: $S$, $D$, $B_S$, and $B'_S$.\\
Eqs.~\eqref{eq:Ps_final}-\eqref{eq:f_new_final}-\eqref{eq:multipl_integr_approx} are the first of the main results presented in this work: a universal multiplicity function for a generic moving barrier, which holds both for haloes and voids from small to very large scales $S$. In Appendix~\ref{ap:local_bias} we discuss the local bias expansion in the effective barrier approach.

\begin{figure}[t!]
\centering
\includegraphics[width=0.67\textwidth]{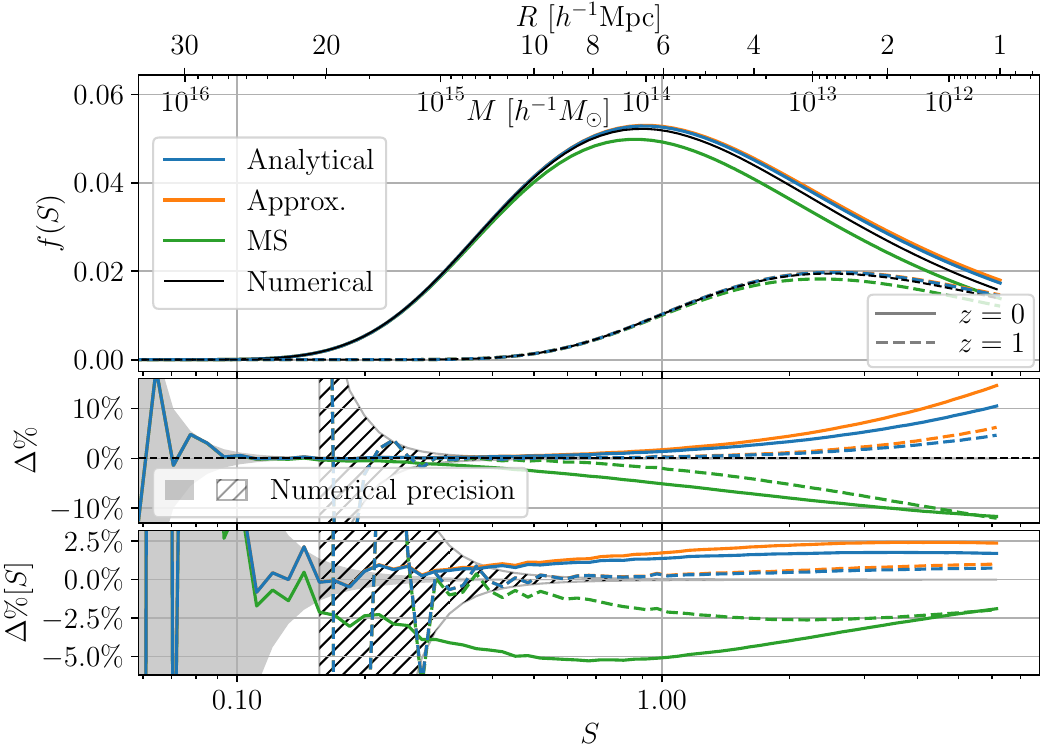}
\caption{Top panel: 
numerical multiplicity function via MC realisations, Eqs.~\eqref{eq:delta_sigle_cholesky}--\eqref{eq:num_f} (black lines); analytical multiplicity function, Eq.~\eqref{eq:f_new_final} (blue lines); corresponding approximation, Eq.~\eqref{eq:multipl_integr_approx} (orange lines); the \citetalias{musso_2012_correlated} multiplicity function (see text, green lines). Solid and dashed lines show $z=0$ and 1 results, respectively. All the multiplicity functions are obtained as a function of $S$ via the \citetalias{sheth_mo_tormen_2001} moving barrier with the top-hat filter. The upper axis shows the corresponding Lagrangian radius and mass.
Middle panel: percent relative differences of the multiplicity functions in the top panel with respect to the numerical multiplicity function, Eqs.~\eqref{eq:delta_sigle_cholesky}--\eqref{eq:num_f}. The shaded and hatched area show the accuracy of the numerical solution.
Bottom panel: 
relative difference between the analytical multiplicity functions, Eqs.~\eqref{eq:f_new_final}--\eqref{eq:multipl_integr_approx} and \citetalias{musso_2012_correlated}, and the numerical one in $S$ units. 
}
\label{fig:multip_analytical}
\end{figure}

The top panel of Fig.~\ref{fig:multip_analytical} shows: the numerical solution of Eqs.~\eqref{eq:langevin_eq_2} via Monte Carlo (MC) realisations, Eqs.~\eqref{eq:delta_sigle_cholesky}--\eqref{eq:num_f} (black lines); the  exact integration of Eq.~\eqref{eq:def_f_new}, i.e. Eq.~\eqref{eq:f_new_final} (blue lines); the corresponding approximation, given by Eq.~\eqref{eq:multipl_integr_approx} (orange lines); the \citetalias{musso_2012_correlated} multiplicity function, with $\Gamma$ appearing in their Eq.~(5) fixed to $1/3$~\citepalias{musso_2012_correlated}, which correspond to $\Gamma_{\delta \delta} = 3/4$ in Eq.~\eqref{eq:multipl_integr_approx} (green lines).  Solid and dashed lines show $z=0$ and 1 results, respectively. All the multiplicity functions are obtained as a function of $S$ via the \citetalias{sheth_mo_tormen_2001} moving barrier with the top-hat filter.
The middle panel shows the percent relative difference of the multiplicity functions in the top panel with respect to the numerical solution. The shaded and hatched areas show the accuracy of the numerical solution for the number of MC realisations used, at $z=0$ and 1 respectively. To better visualise the deviation from the numerical solution, the bottom panel shows the relative  difference between the analytical multiplicity functions, Eqs.~\eqref{eq:f_new_final}--\eqref{eq:multipl_integr_approx} and \citetalias{musso_2012_correlated}, and the numerical one in $S$ units, i.e. the relative difference divided by $S$.
Note the importance of considering the exact $D$ scale dependence: both Eq.~\eqref{eq:f_new_final} (blue) and Eq.~\eqref{eq:multipl_integr_approx} (orange) accurately reproduce the numerical solution on scales considerably smaller than the \citetalias{musso_2012_correlated} model with $\Gamma_{\delta\delta}$ fixed at $3/4$ (green). In particular, at $z=0$ for haloes with mass of $10^{14}$, $10^{13}$, and $5\times 10^{12}~h^{-1}M_\odot$,  Eq.~\eqref{eq:f_new_final} is accurate at $\sim 1 \%$, 3\% and 5\%, while Eq.~\eqref{eq:multipl_integr_approx} is accurate at $\sim1.5 \%$, 5\%, and 7\%, respectively, i.e. roughly $50\%$ less accurate than Eq.~\eqref{eq:f_new_final}. For comparison, the \citetalias{musso_2012_correlated} model ($\Gamma_{\delta \delta}$ fixed) is accurate at $\sim 5\%$, 8.5\%, and 9\%, at the same mass scales. Note, however, that the accuracy largely increases with the redshift: at $z=1$ for the same mass scales Eq.~\eqref{eq:f_new_final} is accurate at $\sim 0.25 \%$, 1\% and 1.5\%, while Eq.~\eqref{eq:multipl_integr_approx} is accurate at $\sim0.25 \%$, 1.5\%, and 2\%, respectively.

\section{The \textit{effective} moving barrier for cosmic voids}\label{sec:effective_barrier}

To compute the moving barrier for haloes and voids we measure it using numerical realisations of 3-dimensional Gaussian random fields, representing the Lagrangian space. The procedure consists in finding haloes and voids, following their definition in Sec.~\ref{sec:ex-set_general}, to measure the corresponding multiplicity function, and to derive the moving barrier generating it (see Appendix~\ref{ap:mcmc_barrier} for details). In this work, we use as Lagrangian space the initial conditions (ICs) of cosmological simulations, which satisfy the above description. 
In this section we focus on voids alone. This is because, in Lagrangian space, the halo treatment is analogous to the void case, but with a positive formation threshold. In addition, in the literature, multiplicity function models for voids are not as advanced as the ones for haloes.

\subsection{Simulations and void finder}
In this work, we use the standard “Dark Energy and Massive Neutrino Universe” (DEMNUni) set of simulations~\citep{carbone_2016_demnuni, parimbelli_2022} and new high-resolution runs with 64 times better mass resolution (HR-DEMNUni)~\citep[see e.g.][]{Beatriz}.
These simulations have been produced using the tree particle mesh-smoothed particle hydrodynamics (TreePM-SPH) code {\sc Gadget-3}~\citep{Springel05}, and are characterised by a Planck 2013~\citep{planck2013} baseline flat \lcdm cosmology with different values of the total neutrino mass and the parameters characterising the dark energy equation of state.
Both the standard and HR runs involve $2048^3$ DM particles and, when present, $2048^3$ neutrino particles, in a volume of $(2000 \, h^{-1}{\rm Mpc})^3$ and $(500 \, h^{-1}{\rm Mpc})^3$, respectively. Here we consider only the massless neutrino flat $\Lambda$CDM case. 
\begin{table}[t!]
\centering
\begin{tabular}{ccccccc}
\toprule
\midrule
$\delta(z_{\rm IC})$ &&  $\delta_{\rm v}$  &&  $\delta^{\rm E}_{\rm v}(z=1)$ &&  $\delta^{\rm E}_{\rm v}(z=0)$  \\
\midrule
-0.035  &&  -2.786  &&  -0.684  &&  -0.800  \\
-0.023  &&  -1.812  &&  -0.566  &&  -0.702  \\
-0.016  &&  -1.253  &&  -0.463  &&  -0.603  \\
-0.011  &&  -0.858  &&  -0.362 &&  -0.497  \\ 
-0.008  &&  -0.623  &&  -0.287  &&  -0.409  \\
-0.0064 &&  -0.497  &&  -0.241  &&  -0.352  \\
-0.005  &&  -0.388  &&  -0.197 &&  -0.294  \\ 
\midrule
\bottomrule
\end{tabular}
\caption{Void formation threshold explored. The first column, $\delta(z_{\rm IC})$, lists the actual density contrast in the simulation ICs at $z_{\rm IC}=99$; the second column, $\delta_{\rm v}$, is the corresponding linear density contrast, linearly evolved to $z=0$; the last two columns, $\delta^{\rm E}_{\rm v}(z=1)$ and $\delta^{\rm E}_{\rm v}(z=0)$, list the corresponding Eulerian (non-linear) density contrast evolved at $z=1, 0$, assuming the spherical symmetry.}
\label{tab:thresholds}
\end{table}
In particular, we consider the initial matter density field of the \lcdm simulation, at $z=99$, for both the DEMNUni and HR-DEMNUni simulations. This set of simulations fits well with the goal of this work, as their ICs are produced at high redshift with the Zeld'ovich approximation, ensuring that the realisation of the field is Gaussian and fully described by the linear theory. Moreover, their combination of volume and resolution ensure a statistically relevant sample of voids from small to large scales~\citep{kreisch_2019,schuster_2019,verza_2019,verza_2022,verza_2023,vielzeuf_2023}.

To identify voids according to the definition given in Sec.~\ref{sec:ex-set_general}, we used the Pylians3\footnote{\url{https://github.com/franciscovillaescusa/Pylians3}}~\citep{Pylians} implementation of the void finding algorithm of~\citep{banerjee_2016}, 
therefore no post-processing of the void catalogue is needed. We explore several void formation thresholds, listed in Tab.~\ref{tab:thresholds}. The first column lists their actual values in the simulation ICs at $z=99$; the second one the corresponding values linearly evolved at $z=0$; the last two columns, $\delta^{\rm E}_{\rm v}(z=1)$ and $\delta^{\rm E}_{\rm v}(z=0)$, list the corresponding Eulerian (non-linear) density contrast evolved down to $z=1$ and 0, assuming the spherical symmetry~\citep[\citetalias{SVdW};][]{pace_2010,pace_2017}.
The exact mapping from the Lagrangian to the Eulerian density contrast may be slightly different from the spherical evolution. Nevertheless, for voids this is an accurate approximation~\citep[][see Sec.~\ref{sec:eulerian_mapping}]{massara2018}. These values are chosen to span a wide range of nonlinear values, from $\delta^{\rm E}_{\rm v}(z=0)\sim -0.8$ to $\delta^{\rm E}_{\rm v}(z=0)\sim -0.3$. Note that the void sample target of galaxy surveys at $z \lesssim 2$ have a DM density contrast in a range between $-0.4$ and $-0.3$~\citep{euclid_vsf}.

\subsection{Methods and results}To find the effective moving barrier for voids, we fit the moving barrier against the multiplicity function derived from the Lagrangian VSF (i.e. measured in the ICs density field of DEMNUni and HR-DEMNUni simulations), considering the general functional shape~\citep[\citetalias{sheth_mo_tormen_2001};][]{,sheth_tormen_2002,paranjape_lam_shet_2012,sheth_2013}
\begin{figure}[t!]
\centering
\includegraphics[width=0.67\textwidth]
{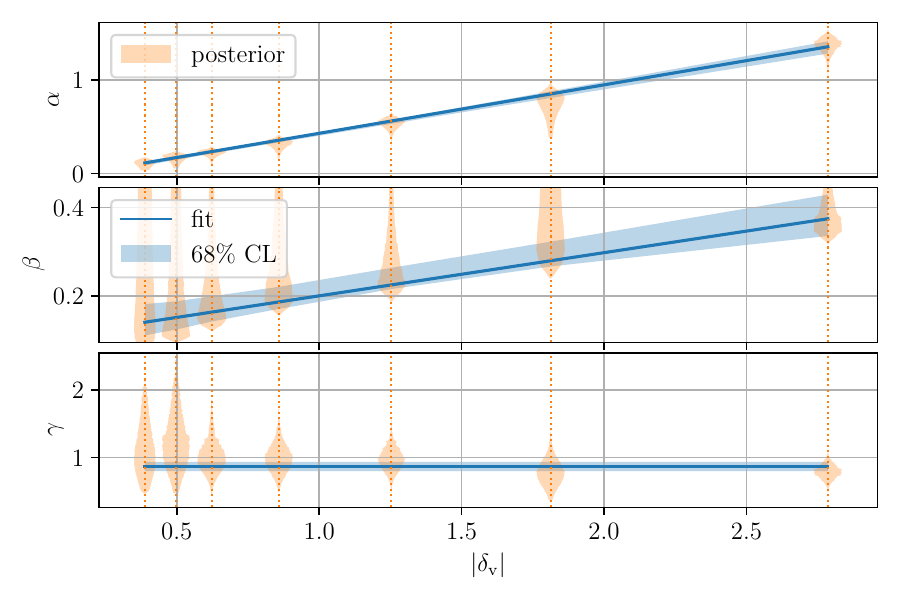}
\caption{
Projected posterior distribution (orange violins) of the $\alpha$, $\beta$, $\gamma$ parameters of Eq.~\eqref{eq:moving_b}, measured against the Lagrangian VSF in the DEMNUni and HR-DEMNUni ICs, as a function of the formation threshold threshold $\delta_{\rm v}$ (vertical orange dashed lines); each panel shows the corresponding projection. Blue solid lines and dashed areas show the best-fit and 68\% CL of the scaling relations described in the text.
}
\label{fig:params_wrt_delta}
\end{figure}
\begin{equation}\label{eq:moving_b}
B(\sigma) = \alpha \left[ 1 + (\beta / \sigma)^{\gamma} \right].
\end{equation}
We perform a Bayesian MCMC analysis of the theoretical multiplicity function corresponding to the above three-parameter moving barrier, to fit $\alpha$, $\beta$, and $\gamma$ as explained in Appendix \ref{ap:mcmc_barrier}.
In the above equation, $\alpha$ modulates the global barrier amplitude, so we expect a dependence from the void formation threshold. The $\beta / \sigma$ quantity has a clear relation with the threshold density in terms of the standard deviation of matter fluctuations, $\sigma = \sqrt{S}$, usually defined as $\nu = |\delta_{\rm v}| / \sigma$. Therefore, a dependence of $\beta$ on $|\delta_{\rm v}|$ is expected. Finally, the $\gamma$ parameter controls the shape of the moving barrier.
Fig.~\ref{fig:params_wrt_delta} shows the projected posterior distribution (orange violins) for the parameters in Eq.~\eqref{eq:moving_b} resulting from each explored void formation threshold, $\delta_{\rm v}$, whose reported value is the one linearly extrapolated at $z=0$. We find that the parameters $\alpha$ and $\beta$ are consistently described by a linear relation in terms of the linear void formation threshold. The $\gamma$ parameter can be consistently described as constant at any formation threshold.  Implementing a linear fit with a MCMC, we obtain
\begin{equation}\label{eq:moving_b_params_val}
\begin{split}
&\begin{matrix}
 \\
\alpha = m_\alpha |\delta_{\rm v}| + q_\alpha:  \\
\beta = m_\beta |\delta_{\rm v}| + q_\beta:  
\end{matrix} \quad \begin{matrix}
m \\
0.517 \pm^{0.031}_{0.035}  \\ 
0.098 \pm^{0.032}_{0.034} 
\end{matrix} \ \left\vert \ \begin{matrix}
q \\
-0.089 \pm^{0.031}_{0.036} \\ 
0.103 \pm^{0.053}_{0.043} 
\end{matrix} \right. \\
&\gamma = 0.87 \pm^{0.07}_{0.07} . 
\end{split}
\end{equation}
These results are shown in Fig.~\ref{fig:params_wrt_delta} as the blue solid lines, while blue shaded areas represent the 68\% credibility level (CL).
\begin{figure}[t!]
\centering
\includegraphics[width=0.95\textwidth]{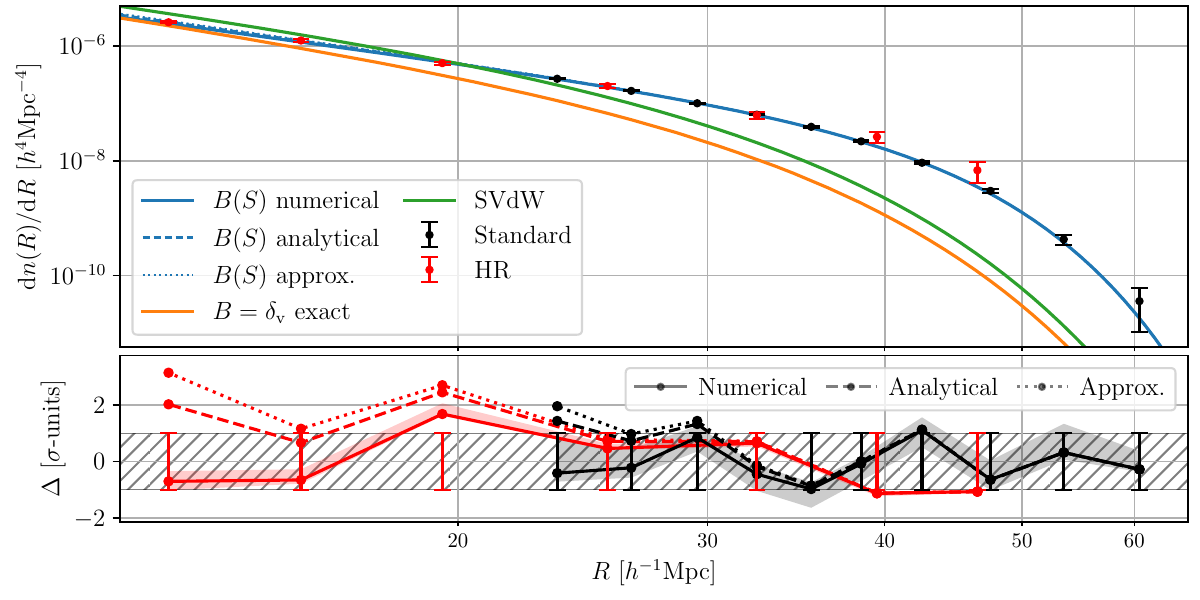}
\caption{Lagrangian VSF for the linear threshold $\delta_{\rm v}=-0.623$. Top panel: measurements from the DEMNUni ICs (black dots), HR-DEMNUni ICs (red dots) with the corresponding Poissonian uncertainty (error bars); VSF from the effective barrier approach presented in this work, Eq.~\eqref{eq:moving_b}, with the MCMC best-fit parameters, Eq.~\eqref{eq:moving_b_params_val}. The solid blue line shows the exact numerical solution, the blue dashed line the analytical multiplicity function, Eq.~\eqref{eq:f_new_final}, and the blue dotted line the corresponding approximation, Eq.~\eqref{eq:multipl_integr_approx};
VSF from the standard excursion-set approach with a top-hat filter, i.e. $B=\delta_{\rm v}$ in Eqs.~\eqref{eq:delta_sigle_cholesky}--\eqref{eq:num_f} (orange line); 
\citetalias{SVdW} model (green line).
Bottom panel: residuals of the VSF obtained via the effective barrier best-fit with respect to the VSF measured in the DEMNUni ICs (black lines) and HR-DEMNUni ICs (red lines), in units of the Poissonian uncertainty. Solid lines shows the exact numerical solution, dashed lines the analytical multiplicity function, Eq.~\eqref{eq:f_new_final}, and dotted line the corresponding approximation, Eq.~\eqref{eq:multipl_integr_approx}.}
\label{fig:VSF_plot}
\end{figure}

The upper panel of Fig.~\ref{fig:VSF_plot} shows the Lagrangian VSF with linear formation threshold $\delta_{\rm v}=-0.623$, 
as measured from the DEMNUni ICs (black dots)  and HR-DEMNUni ICs (red dots), with the corresponding Poissonian uncertanty (errorbars), together with the following theoretical modellings:
i) the VSF from the standard excursion-set multiplicity function, i.e. using a top-hat filter and a constant void formation threshold corresponding to the physical one, $B(S)=\delta_{\rm v}$ in Eqs.~\eqref{eq:delta_sigle_cholesky}--\eqref{eq:num_f} (orange curve);
ii) Lagrangian VSF of \citetalias{SVdW} (green curve) with the corresponding $\delta_{\rm v}=-0.623$: 
the global behaviour is similar to the previous case, but its amplitude is slightly higher. This is because the \citetalias{SVdW} multiplicity function is the exact first crossing double barrier solution with the sharp-$k$ filter, i.e. of Markovian random walks, for which the crossing probability is higher;
iii) the VSF from the effective barrier approach presented in this work, Eq.~\eqref{eq:moving_b}, with the MCMC best-fit parameters, Eq.~\eqref{eq:moving_b_params_val}. The solid blue line shows the exact numerical solution, Eqs.~\eqref{eq:delta_sigle_cholesky}--\eqref{eq:num_f}, the blue dashed line the analytical multiplicity function, Eq.~\eqref{eq:f_new_final}, and the blue dotted line the corresponding approximation, Eq.~\eqref{eq:multipl_integr_approx}. 
The shaded area shows the 68\% CL interval of the posterior distribution of the parameters propagated to the VSF. Finally, the lower panel of Fig.~\ref{fig:VSF_plot} shows the residuals of the VSF obtained via our new multiplicity function and the effective barrier model with respect to VSF measurements from the DEMNUni ICs (black lines) and HR-DEMNUni ICS (red lines), in units of the Poissonian uncertainty. The solid lines shows the exact numerical solution, Eqs.~\eqref{eq:delta_sigle_cholesky}--\eqref{eq:num_f}, dashed lines the analytical multiplicity function, Eq.~\eqref{eq:f_new_final}, and blue dotted line the corresponding approximation, Eq.~\eqref{eq:multipl_integr_approx}. 
The shaded areas show the 68\% CL, while the hatched area shows $\pm 1 \sigma$ interval. The agreement is within $1$-$\sigma$ on almost all the scales. 
The threshold $\delta_{\rm v}=-0.623$ is chosen as it is representative of the void population detectable around $z\sim 1$ in galaxy surveys~\citep{euclid_vsf}. The results of the other thresholds explored are qualitatively analogous. 
Eqs.~\eqref{eq:moving_b}--\eqref{eq:moving_b_params_val} represent the second most important result of our work.

As final consideration here we have modelled the void multiplicity function with a single moving barrier. This may seam in contrast to what written in Sec.~\ref{sec:ex-set_general}, i.e. that void formation is characterised by two barriers: the formation and the collapsing ones. The double barrier was introduced to account for the void-in-cloud effect, which however occurs at scales smaller than the ones explored in this work~\citepalias{SVdW}. Moreover, the use of a top-hat filter and the focus on minima in the density field reasonably push the void-in-cloud effect at scales even smaller than in \citetalias{SVdW}\footnote{This last condition describes the filtered density fields as Markovian random walks corresponding to a large variation of the field at relatively high $\sigma$ (low $R$). This enhances the void-in-cloud fraction with respect to the correlated step case.}. 
We verified this by measuring in the DEMNUni-ICs the density profile of each void up to 100~$h^{-1}$Mpc. We have not found any void embedded in large overdensities that reaches neither the linear spherical collapsing threshold extrapolated at $z=0$, $\delta_c \simeq 1.686$, nor the turn-around threshold,  $\delta_{\rm ta} \simeq 1.06$.

\section{Eulerian mapping}\label{sec:eulerian_mapping}

In this Section, we introduce the Lagrangian to Eulerian mapping assuming the simplest possible approximation: the spherical map. With this assumption, each Lagrangian halo and void is considered evolving following the spherically symmetric gravitational collapse/expansion \citep{SVdW,pace_2010,pace_2017}. For both the halo and the void cases, the number of objects is assumed to be conserved from Lagrangian to Eulerian space. Considering the void case, due to mass conservation, the corresponding Eulerian radius is $R_{\rm E} = (1+\delta_{\rm v}^{\rm E})^{-1/3} R$, where the super- and sub-script E denotes Eulerian quantities. For haloes, the spherical symmetric collapse implies that all fluid elements of a Lagrangian halo collapse in the Eulerian space, therefore $M_{\rm E}(R) = M_{\rm TH}(R)$ as defined in Sec.~\ref{sec:ex-set_general}. Note that the multiplicity function is a differential quantity; therefore, to express it as a function of Eulerian quantities, we use the relation
\begin{equation}
f(S) \, {\rm d} S = f(R) \, {\rm d} R = f(R_{\rm E}) \, {\rm d} R_{\rm E}  = f(M_{\rm E}) \, {\rm d} M_{\rm E} .
\end{equation}

\begin{figure}[t!]
\centering
\includegraphics[width=0.95\textwidth]{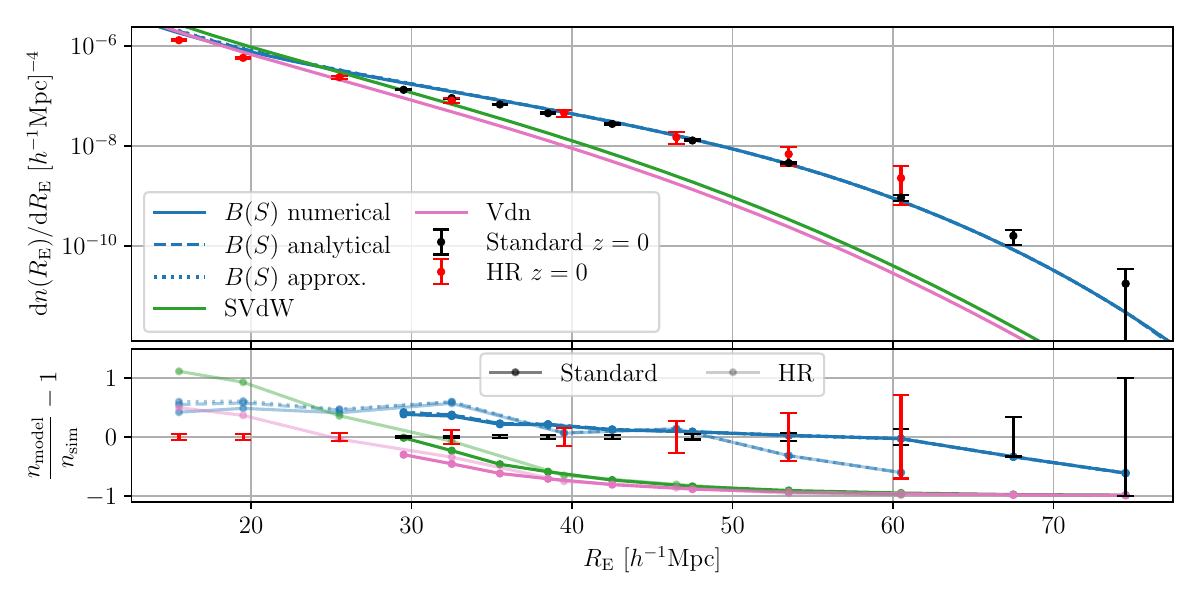}
\caption{Top panel: VSF measurements from DEMNUni (black dots) and HR-DEMNUni (red dots) at $z=0$, for voids reaching $\delta^{\rm E}_{\rm v}=-0.409$ threshold, corresponding to the linear threshold $\delta_{\rm v}=-0.623$ when the spherical symmetric evolution is assumed. Blue curves are the theoretical VSF in Eulerian space, Eq.~\eqref{eq:eulerian_VSF}, obtained assuming: spherical symmetry in void evolution, using the effective moving barrier Eqs.~\eqref{eq:moving_b}--\eqref{eq:moving_b_params_val} for $\delta_{\rm v}=-0.623$, with the corresponding numerical multiplicity function (blue solid); Eqs.~\eqref{eq:delta_sigle_cholesky}--\eqref{eq:num_f}, and the two analytical approximations Eq.~\eqref{eq:f_new_final} (dashed blue) and Eq.~\eqref{eq:multipl_integr_approx} (dotted blue). Eulerian \citetalias{SVdW} model (solid green) and \citetalias{jennings2013} model (solid pink).
Bottom panel: relative difference between the theoretical models appearing in the upper panel (same colours) with respect to the VSF measured in DEMNUni (dark colours) and HR-DEMNUni (transparent). The errorbars show the corresponding Poissonian uncertainty.}
\label{fig:VSF_EU}
\end{figure}

\subsection{The cosmic void case} From the above equation it follows that the Eulerian multiplicity function is $f_{\rm E}(R_{\rm E}) = f[R(R_{\rm E})] \, {\rm d} R / {\rm d} R_{\rm E} = (1+\delta_{\rm v}^{\rm E})^{1/3}f[R(R_{\rm E})]$. 
The Eulerian void size function, therefore, is
\begin{equation}\label{eq:eulerian_VSF}
\frac{{\rm d} n(R_{\rm E})}{{\rm d}R_{\rm E}} = \frac{3}{4 \pi R^3(R_{\rm E})} f_{\rm E}(R_{\rm E}).
\end{equation}
The upper panel of Fig.~\ref{fig:VSF_EU} shows the VSF of voids reaching at $z=0$ the non-linear formation threshold $\delta^{\rm E}_{\rm v} = -0.409$, which corresponds to the spherical non-linear evolution of the linear void formation threshold $\delta_{\rm v} = -0.623$ (see Tab.~\ref{tab:thresholds}). Black and red dots show the VSF for voids with $\delta^{\rm E}_{\rm v} = -0.409$ measured at $z=0$ in the DEMNUni and HR-DEMNUni simulations, respectively. Errorbars show the Poissonian uncertainty. The scales represented correspond to the Lagrangian ones in Fig.~\ref{fig:VSF_plot}, that appear larger in Eulerian space due to void expansion.
The curves show the following theoretical models: \\ 
i) Our VSF model with the effective moving barrier, Eq.~\eqref{eq:moving_b}, using the exact numerical multiplicity function (solid blue) from Eqs.~\eqref{eq:delta_sigle_cholesky}--\eqref{eq:num_f} and the moving barrier of Eq.~\eqref{eq:moving_b} with the MCMC best-fit parameters obtained in Lagrangian space, Eq.~\eqref{eq:moving_b_params_val}. 
The blue dashed line shows the corresponding analytical approximation, Eq.~\eqref{eq:f_new_final}, while the dotted blue line represents the approximated integration, Eq.~\eqref{eq:multipl_integr_approx}. 
They are mapped in Eulerian space assuming a spherical symmetric evolution, Eq.~\eqref{eq:eulerian_VSF}.\\
ii) The \citetalias{SVdW} model (green curve) with linear formation threshold $\delta_{\rm v}=-0.623$, mapped in Eulerian space via Eq.~\eqref{eq:eulerian_VSF}.\\
iii) The Jennings at al. 2013~\citep{jennings2013} VSF model (hereafter~\citetalias{jennings2013}) with linear formation threshold $\delta_{\rm v}=-0.623$ (pink curve). This model, also known as the volume-conserving model, is a Eulerian mapping of the \citetalias{SVdW}, which relies mainly on one assumption: the total volume fraction of voids is conserved from Lagrangian to Eulerian space, at the expense of the conservation of the number of voids. This assumption was introduced to avoid normalisation problems in the sharp-$k$ multiplicity function \citepalias{SVdW} mapped in Eulerian space assuming spherical symmetric evolution. 
\begin{figure}[t!]
\centering
\includegraphics[width=0.95\textwidth]{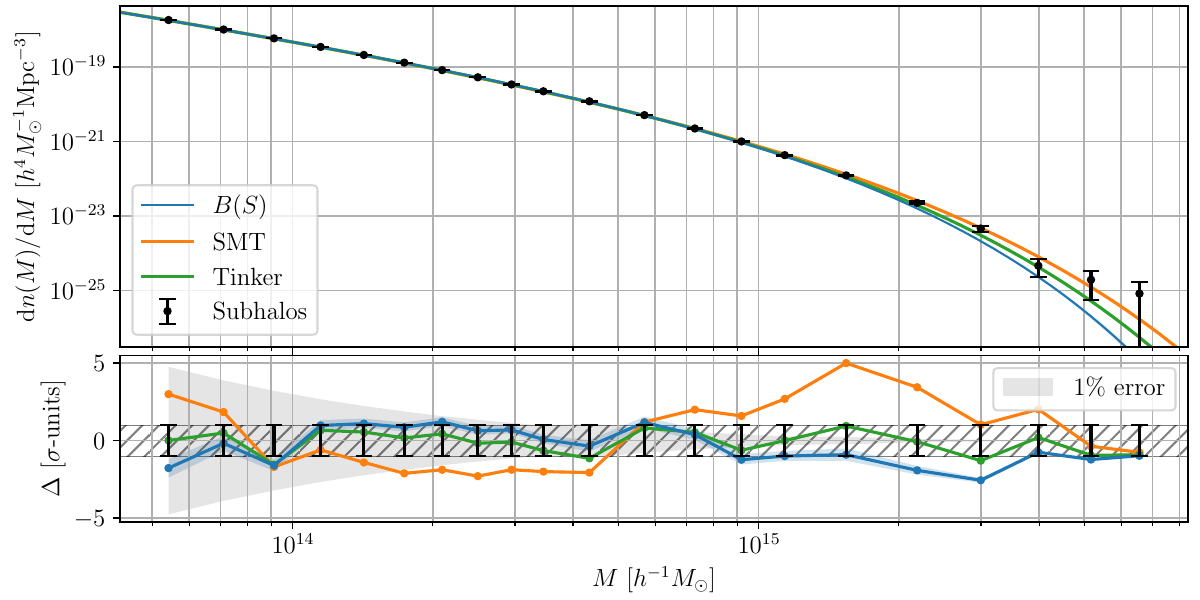}
\caption{Eulerian HMF. Top panel: HMF measured from DEMNUni at $z=0$ (black dots) and corresponding Poissonian uncertainty (errorbars); Eulerian spherical mapping of the effective barrier approach implemented in this work (blue), Eqs.~\eqref{eq:multipl_integr_approx},~\eqref{eq:moving_b},  fitting the halo threshold from Eq.~\eqref{eq:moving_b_params_val}; \citetalias{sheth_mo_tormen_2001} best-fit model (orange); 4-parameters Tinker et al. 2008~\citep{tinker_2008} best-fit model.
Bottom panel: relative difference between the theoretical models of the upper panel (same colour libelling) with respect to HMF measured in the DEMNUni in unit of Poissonian uncertainty. Errorbars and hatched area show $\pm 1\sigma$ interval, dashed area shows the $\pm 1\%$ uncertainty}
\label{fig:HMF_EU}
\end{figure}
\\
\noindent Finally, the lower panel of Fig.~\ref{fig:VSF_EU} shows the residuals at $z=0$ of the above Eulerian VSF models with respect to the VSF measured from DEMNUni (dark colours) and HR-DEMNUni (transparent). Errorbars represent the Poissonian uncertainty. It can be observed that the Eulerian mapping of \citetalias{SVdW} and the \citetalias{jennings2013} model is not able to reproduce the VSF from simulations. Concerning our model, it can be noticed that the spherical mapping is accurate within $1\sigma$ at both large and intermediate scales, while it becomes less accurate at lower scales, on which the spherical Eulerian map over-predicts the number of voids by about $\sim 50\%$. The accuracy at larger scales is in agreement with the fact that the anisotropic part of the strain of the displacement field around density extremants decreases as the scale increases~\citep{bardeen_1986,bond_1996}. In particular, the expected value of the ellipticity is proportional to $\sigma(R)/|\delta_{\rm v}|$~\citepalias{sheth_mo_tormen_2001}, which makes the spherical approximation accurate on large scales. On smaller scales, other effects, such as non-sphericity and non-locality, may become relevant.
From the same arguments, it follows that the accuracy of the spherical Eulerian mapping increases with the redshift. \\

\subsection{The DM halo case} Lagrangian haloes undergo a full collapse in Eulerian space, which makes the mapping conceptually different from the void case. On the one hand, at each redshift it corresponds a unique linear collapse threshold (see Appendix~\ref{ap:redshift_evolution} for details); on the other hand, the spherical symmetric evolution can hardly be accurate in the strongly nonlinear regime of the collapse. 
In the literature it is shown that, at the first-order approximation, the non-spherical collapse is reflected in a scale-dependent~\citepalias{sheth_mo_tormen_2001} and fuzzy formation threshold~\citep{robertson_2009,ludlow_porciani_2011} in Lagrangian space. In addition, non-linear evolution can break the top-hat Lagrangian to Eulerian mass conservation, $M_{\rm E}(R) = M_{\rm TH}(R)$, as the fluid elements of a Lagrangian spherical overdensity may not all and exclusively collapse in the corresponding Eulerian halo~\citep{ludlow_porciani_2011}. Finally, above all, there is the problem of estimating which regions in the Universe (or in simulations) are collapsed and eventually virialized, i.e. are DM haloes. Different prescriptions result in different halo finders~\citep{knebe_2013_halo_finders}. This makes an {\it a priori} modelling of haloes statistics, such as the HMF, extremely challenging. As a consequence, the halo multiplicity function needs to be fitted according to the different halo finders considered\footnote{Note that analogous considerations are valid for void finding algorithms for which the corresponding void definition cannot be easily mapped to the excursion-peak theory one. In that case, exactly as for haloes, the effective barrier describing the statistics needs to be fitted against simulations.}~[\citealp{tinker_2008},~\citetalias{sheth_mo_tormen_2001}], and for each different halo finding algorithm, a different effective barrier would correspond to it, representing the statistics of each different halo population.
In this perspective, we use the HMF measured at $z=0$ from the DEMNUni simulations to find the corresponding halo formation threshold for Lagrangian haloes, as defined in Sec.~\ref{sec:ex-set_general}. In particular, to test the stability of our approach against the used halo-finder, we consider different halo catalogues, such as: Friend-of-Friends (FoF) haloes; subhaloes obtained by post-processing FoF haloes with the {\sc  Gadget-3 Subfind} algorithm (which identifies locally overdense regions, i.e. areas enclosed by an isodensity contour that traverses a saddle point)~\citep{springel_2001_gadeget,dolang_2009_gadget}; and Spherical-Overdensity (SO) halo catalogues corresponding to an overdensity threshold equal to the virial value and obtained by exploiting the SO package within the {\sc Subfind} algorithm. We found that the barrier slightly changes to accommodate the HMFs of the different catalogues, while variations on the formation threshold values are minor.

Then, according to the spherical mapping approximation, we assume the linear halo formation threshold, $\delta_{\rm h}$, to be scale independent and the Eulerian mass to be the same as the Lagrangian one, $M_{\rm E}(R) = M_{\rm TH}(R)$.  
Following~\citetalias{sheth_mo_tormen_2001}, we consider the subhalo catalogues and, as explained in Appendix \ref{ap:redshift_evolution}, we use Eq.~\eqref{eq:moving_b_params_val} as a prior to fit the threshold, after substituting $\delta_{\rm v}$ with $\delta_{\rm h}$. We find a redshift independent best-fit $\delta_{\rm h} = 2.424 \pm^{0.076}_{0.071}$. Moreover the corresponding effective barrier parameters in Eq.~\eqref{eq:moving_b} are found to be redshift independent as well, satisfying the regime in which the HMF can be considered self-similar, confirming its redshift universality (see Appendix \ref{ap:redshift_evolution} for details). Note that, even if this value seems to be larger than the usual spherical collapse threshold, $\delta_{\rm c} \simeq 1.686$, actually our results confirm previous finding in the literature~\citep{robertson_2009}. In fact, the barrier height corresponding to $\delta_{\rm h} = 2.424 \pm^{0.076}_{0.071}$ is $\alpha = 1.209 \pm^{0.008}_{0.007}$, only a $\sim 15\% $ different from the ~\citetalias
{sheth_mo_tormen_2001} moving barrier, $\sqrt{0.7} \delta_{\rm c}\sim 1.4$.

Fig.~\ref{fig:HMF_EU} shows the results. The upper panel shows: i) the sub-HMF measured at $z=0$ from DEMNUni (black dots), with the corresponding Poissonian uncertainty (error bars); ii) threshold fit $\delta_{\rm h}$ of theoretical HMF from the moving barrier approach using Eqs.~\eqref{eq:multipl_integr_approx}--\eqref{eq:moving_b}--\eqref{eq:moving_b_params_val} (blue curve); iii) the \citetalias{sheth_mo_tormen_2001} best-fit model (orange curve); iv) 4-parameters Tinker et al. 2008~\citep{tinker_2008} best-fit model.
The bottom panel shows the relative differences between the measurements from simulations and the best-fit models of the upper panel in units of the Poissonian uncertainty. The errorbars and the hatched area show the $\pm 1\sigma$ interval, the dashed area shows the $\pm 1\%$ uncertainty. It can be noticed that, on the scales considered, the model presented in this work, Eqs.~\eqref{eq:multipl_integr_approx}--\eqref{eq:moving_b}--\eqref{eq:moving_b_params_val}, plus the fitting of the linear formation threshold $\delta_{\rm h}$, allow us to reach a better accuracy than for the \citetalias{sheth_mo_tormen_2001} best-fit model and similar to the 4-parameters Tinker et al. 2008~\citep{tinker_2008} best-fit model.
Appendix~\ref{ap:redshift_evolution} shows the redshift dependence of all these theoretical HMFs.

A more accurate Lagrangian to Eulerian mapping should get rid of the spherical approximation, and include off-diagonal terms of the strain tensor as well as tidal effects. This is beyond the scope of this work, which focuses on Lagrangian space, and we leave it for future studies.

\section{The void density profile from the effective barrier}\label{sec:profile}

In this Section, we explore the effective barrier approach for statistics beyond the multiplicity function, by considering the void density profile of Lagrangian voids. Note that this is the integral of the void-matter cross-correlation function. 
Here we focus on the density profile of voids because, beyond the motivations explained at the beginning of Sec.~\ref{sec:effective_barrier}, halo collapse destroys the one-to-one map between the Lagrangian and Eulerian profile, which is not the case for voids. 
Moreover, a theoretical modelling of the Lagrangian density profile of voids would be useful for their analysis in the redshift space~\citep{pisani_2014,hawken_2017,hamaus_2016,cai_2016,radinovic_2023}.

In the standard excursion-set approach, each stochastic realisation of the $\delta({\bf q},R)$ field, given by Eqs.~\eqref{eq:langevin_eq_1}, statistically corresponds to a realisation of the integrated density profile with respect to a random Lagrangian position ${\bf q}$. Here, we apply this feature using our effective barrier approach to void formation, in order to verify whether it can reproduce the density profile around minima. 
As the effective barrier $B$ in Eqs.~\eqref{eq:moving_b}--\eqref{eq:moving_b_params_val} does not correspond to the void formation threshold $\delta_{\rm v}$, the corresponding density profile realisations would not describe the physical ones.  
We avoid this effect by considering two steps. 
First, at scales larger than the crossing radius, $R_\times$, we normalise the field to the actual void formation threshold as
\begin{equation}\label{eq:profile_1}
\delta(R) \rightarrow \delta(R) \,  \delta_{\rm v} / B(R) \quad {\rm for} \quad R \geq R_\times .
\end{equation}
This step aims to rescale the amplitude of the field from the effective to the physical one. 
Second, at scales smaller than the crossing radius, we shift the stochastic realisation of the field as
\begin{equation}\label{eq:profile_2}
\delta(R) \rightarrow \delta(R) - B(R_\times) + \delta_{\rm v} \quad {\rm for} \quad R < R_\times .
\end{equation}
This shifts the stochastic realisation $\delta(R)$ by an offset chosen to match the physical void formation threshold at the crossing radius, $\delta(R_\times) = \delta_{\rm v}$. After crossing, the field behaves as a random walk pinned at the coordinate $(R_\times,\delta_{\rm v})$ in the $R-\delta$ plane and therefore satisfies the above equation. 
It follows that the width of the distribution of void profiles is approximated by $\sqrt{s-S(R_\times)}$ as $s$ increases ($R$ decreases).

\begin{figure}[t!]
\centering
\includegraphics[width=0.95\textwidth]{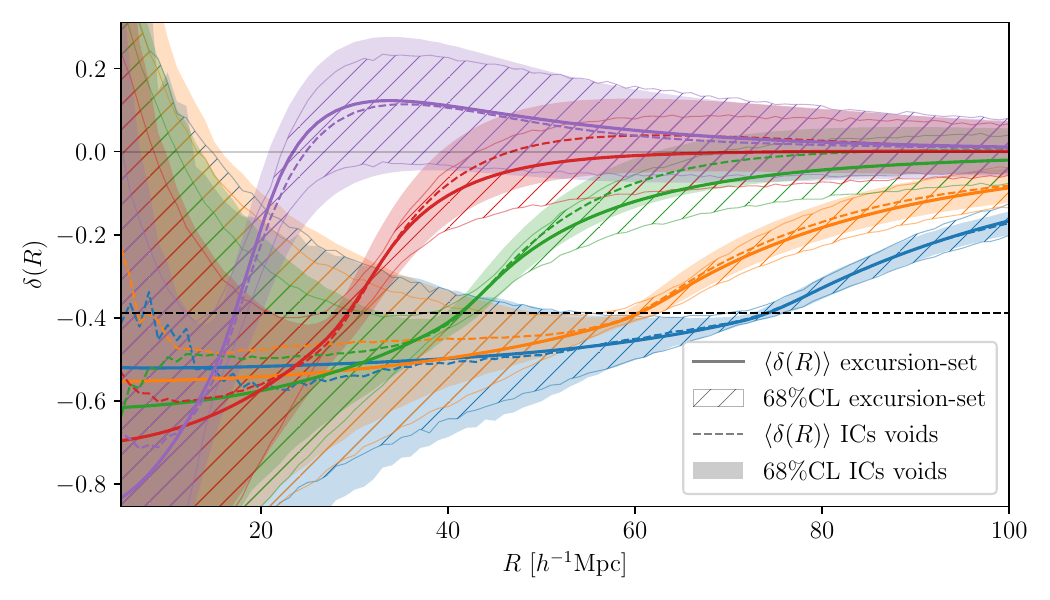}
\caption{Lagrangian density profile, linearly evolved at $z=0$, for voids with linear formation threshold of $\delta_{\rm v}=-0.388$ and void radius between $15-21$Mpc (purple), $33-39$Mpc (red), $58-65$Mpc (green), and $(72-79)$~$h^{-1}$Mpc (blue). Solid lines show the excursion-set prediction with the effective moving barrier obtained as described in \ref{sec:profile}; light hatched areas the shortest interval subtending 68\% of the distribution; dashed lines show the mean void density profile measured in simulations ICs; light shaded areas the shortest interval subtending 68\% of the distribution.
}
\label{fig:profile}
\end{figure}

Fig.~\ref{fig:profile} shows the comparison between the void density profile measured in simulations ICs and linearly evolved at $z=0$ (dashed lines), with respect to stochastic realisations of $\delta(R)$, obtained with the corresponding effective moving barrier, Eqs.~\eqref{eq:moving_b}--\eqref{eq:moving_b_params_val}, rescaled via Eqs.~\eqref{eq:profile_1}--\eqref{eq:profile_2} (solid lines). Different colours represent voids with different formation radii, as specified in the caption. 
The shaded areas show the shortest interval subtending $68\%$ of the distribution of density profiles in simulations ICs; while the hatched areas show the same quantity for stochastic realisations.
Interestingly, in both cases, the distributions are well described by a Gaussian PDF for each radius $R$.
The correspondence between simulations ICs and stochastic realisations is very good, in particular: i) the void profile distribution obtained via the effective barrier approach closely matches the one from simulations ICs, as can be noted by the shaded and hatched areas in Fig.~\ref{fig:profile}; ii) the mean of the excursion-set profile distribution reproduces the global behaviour of the mean profile measured in the simulations ICs, while the accuracy depends on the void scale considered; iii) the change of slope around the crossing radius is accurately reproduced; iv) interestingly, the effective barrier predicts the formation of a non-collapsing overdensity in which smaller voids are embedded, as observed in simulations ICs. 

Our void profile prediction in Lagrangian space is the third main result of this work, and the methodology to model it has been presented here for the first time in the literature. In particular, this modelling uses the same effective barrier of Eqs.~\eqref{eq:moving_b}--\eqref{eq:moving_b_params_val} used to derive the theoretical multiplicity function and VSF presented in this work. This modelling would not be representative of simulation ICs measurements unless the effective barrier contained statistical information beyond the multiplicity function. In this work we do not explore the Lagrangian to Eulerian mapping for the void density profile, which we aim to study in a future work. However, it has already been shown in the literature that it is possible to obtain the Eulerian void density profile from the Lagrangian one analytically ~\citep{massara2018,stopyra_2021} and we think that similar techniques can be applied also to our formalism.

\section{Conclusions}\label{sec:conclu}
In this work, we show that using an effective moving barrier in the excursion-set framework, it is possible to combine the latter with the peak theory in Lagrangian space. In particular: 
i) we provide in Sec.~\ref{sec:ex-set_general} a precise halo and void formation definition in Lagrangian space, that was missing for voids, which combines excursion-set and peak theory; 
ii) in Sec.~\ref{sec:multipl_func}, for a general moving barrier in the presence of scale correlations, we derive Eqs.~\eqref{eq:f_new_final} and~\eqref{eq:multipl_integr_approx}, i.e. a multiplicity function able to recover both halo and void statistics from small to very large scales $S$. 
In particular, we show that the exact computation of $\langle \delta'^2_S \rangle$ is crucial to obtain accurate analytical approximations of the general solution (see Fig.~\ref{fig:multip_analytical} for a detailed comparison with the numerical multiplicity function obtained via MC realisations);
iii) in Sec.~\ref{sec:effective_barrier} we derive, in the void case, the parameters characterising the \citetalias{sheth_mo_tormen_2001}-like effective moving barrier, in Eq.~\eqref{eq:moving_b},  as a function of the void formation threshold, $\delta_{\rm v}$, by inserting Eq.~\eqref{eq:moving_b} in Eq.~\eqref{eq:num_f}, Eq.~\eqref{eq:f_new_final} and Eq.~\eqref{eq:multipl_integr_approx}, and fitting the VSF obtained via the latter against the Lagrangian VSF measured in the DEMNUni ICs; 
iv) in Sec.~\ref{sec:eulerian_mapping} we discuss the Lagrangian to Eulerian mapping, in particular we implement the spherical map in Eulerian space for the Lagrangian VSF and HMF, introduced in Secs.~\ref{sec:multipl_func}--\ref{sec:effective_barrier}. We obtain a good agreement with simulations for intermediate and large voids and we accurately reproduce the HMF by fitting the linear halo formation threshold from simulations; v) in Sec.~\ref{sec:profile} we show that, in the excursion-set framework with the effective barrier, it is possible to derive the Lagrangian void density profile, using the effective barrier of Eqs.~\eqref{eq:moving_b}--\eqref{eq:moving_b_params_val} entering the corresponding multiplicity function, Eq.~\eqref{eq:num_f}, derived in this work. See Fig.~\ref{fig:profile} for a comparison against the DEMNUni void profiles.

Our results may have several applications in upcoming galaxy surveys analysis, in particular concerning DM tracer statistics as the HMF, VSF, and the void-galaxy cross-correlation. As shown in Fig.~\ref{fig:VSF_plot}, our multiplicity function provides a much better accuracy than previous modellings when compared with measurements from cosmological simulations.  
Moreover, it is accurate up and beyond the scale of the lightest clusters~\citep{Adam_2019_Euclid_clusters,hilton_2021_clusters_ACT} and the smallest voids detectable in galaxy surveys~\citep{contarini_2021}. 
In addition, an accurate modelling of the Lagrangian void density profile can help improving the modelling of redshift space distortions around voids~\citep{pisani_2014,hamaus_2014_vg_xc,hamaus_2016,cai_2016,nadathur_2019,nadathur_2019_BOSS,hamaus_2020,nadathur_2020,hamaus_2022_euclid,woodfinden_nadathur_2022, radinovic_2023}. These results are particularly relevant for data analyses from ongoing and forthcoming galaxy redshift surveys, such as Euclid~\citep{laureijs_20211_euclid_report}, Roman~\citep{spergel_2015_WFIRST}, SPHEREx~\citep{dore_2018_SPHEREx}, DESI~\citep{DESI_2016}, PFS~\citep{PFS_2016}, etc., where the amount of data and the galaxy density will require high precision and accuracy in the theoretical modelling, from the smaller to the larger scales.
In this respect, note that in this work we considered around $2\times10^5$ voids (see Tab~\ref{tab:Rmin}), while, for example, in Euclid the number of expected voids is around 6000~\citep{euclid_vsf}. This makes the modelling presented in this work very accurate, with much smaller uncertainties than those expected from future surveys.
In this respect, in a future work we plan to further explore the mapping from the Lagrangian to the Eulerian space at scales smaller than $\sim 40 h^{-1}{\rm Mpc}$ at $z=0$, on which the spherical mapping already holds within $1\sigma$. For voids in galaxy surveys other observational effects beyond the Lagrangian to Eulerian mapping in real space should be considered, such as the redshift-space distortion impact. We plan to explore them in future works by applying well known techniques in the literature~\citep[e.g.][]{pisani_2015_rsd,hamaus_2015,cai_2016,nadathur_2019, correa_2019,correa_2021}.

\vspace{1.5em}

\noindent  
GV would like to thank Nickolas Kokron for useful discussions.
GV acknowledges NASA grant EUCLID12-0004. GV and AP acknowledge support from the Simons Foundation to the Center for Computational Astrophysics at the Flatiron Institute.
The DEMNUni simulations were carried out in the framework of ``The Dark Energy and Massive-Neutrino Universe" project, using the Tier-0 IBM BG/Q Fermi machine and the Tier-0 Intel OmniPath Cluster Marconi-A1 of the Centro Interuniversitario del Nord-Est per il Calcolo Elettronico (CINECA). We acknowledge a generous CPU and storage allocation by the Italian Super-Computing Resource Allocation (ISCRA) as well as from the coordination of the ``Accordo Quadro MoU per lo svolgimento di attività congiunta di ricerca Nuove frontiere in Astrofisica: HPC e Data Exploration di nuova generazione'', together with storage from INFN-CNAF and INAF-IA2.

\bibliographystyle{JHEP}
\bibliography{biblio}

\appendix

\section{Correlation of the density contrast field scale derivatives}\label{ap:compute_W}
The explicit expression of $\langle \delta'^2_S \rangle = \langle ({\rm d} \delta_S / {\rm d} S)^2 \rangle$ for a generic filter function $W(kR)$ reads
\begin{equation}
\langle \delta'^2_S \rangle = \left( \frac{{\rm d}R}{{\rm d} S} \right)^2 \left\langle \left( \frac{\partial \delta(R)}{\partial R} \right)^2 \right\rangle ,
\end{equation}
where, considering Eqs.~\eqref{eq:filtered_delta}--\eqref{eq:langevin_eq_1},
\begin{equation}\label{eq:brute_force_int}
\left\langle \left( \frac{\partial \delta(R)}{\partial R} \right)^2 \right\rangle = \int_0^\infty \frac{\mathrm{d}k \, k^2}{2 \pi^2} P(k) \left( \frac{\partial W(k R)}{\partial R} \right)^2.
\end{equation}
The derivative of the top-hat filter function with respect to the smoothing lenght $R$ is
\begin{equation}
\frac{\partial W(k R)}{\partial R} = \frac{3}{R x^3} \left[\left(x^2 - 3\right) \sin (x) + 3 x \cos(x) \right]\,,
\end{equation}
with $x = kR$.
Note that for $k \gg 1/R$, this function behaves as $\sim \sin(kR) / k$. Therefore, for large $k$, the integrand of Eq.~\eqref{eq:brute_force_int} behaves as $\sim P(k) \sin^2(kR)$. As a result, the convergence of the integral relies solely on the behaviour of $P(k)$ at large $k$, without any additional suppression term; therefore, in the cases where the integral convergences, the convergence is slow~\citep[see Appendix B in][]{paranjape_lam_shet_2012}. This makes the direct evaluation of this integral computationally expensive, as it requires the knowledge of $P(k)$ up to very high $k$ values. As a consequence, in the literature, this quantity is generally not computed with a top-hat filter; even when the top-hat filter is used for other quantities, such as $\langle \delta^2(R) \rangle$. Usually, $\langle \delta'^2_S \rangle$ is evaluated with a Gaussian filter smoothed at $R_{\rm G}$, and mapped to a different top-hat scale $R_{\rm TH}$ according to various prescriptions~\citep{porciani_matarrese1998,paranjape_2013,paranjape_2013_bias,biagetti_2014}.

\begin{figure}[t!]
\centering
\includegraphics[width=0.67\textwidth]{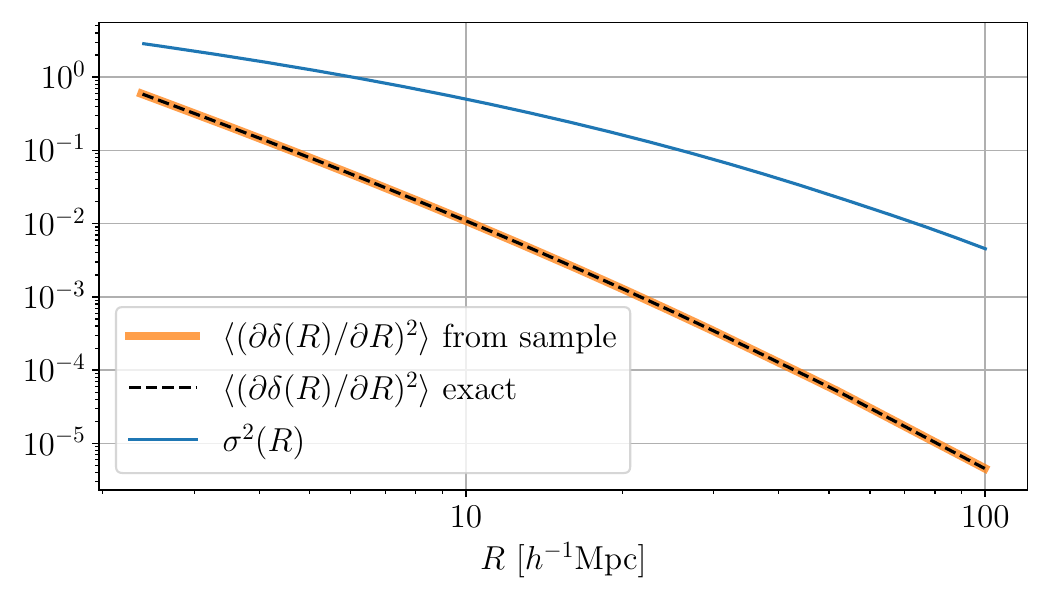}
\caption{$\langle (\partial \delta(R) / \partial R)^2 \rangle$ computed as the ensemble variance of MC realisations (orange line); same quantity computed with Eq.~\eqref{eq:der_variance} (black dashed line); $\sigma^2(R)=\langle \delta^2(R) \rangle$ (blue line). All the quantity showed are computed with a top-hat filter}
\label{fig:variance_derivative}
\end{figure}

We propose here a different approach to exactly compute $\langle \delta'^2_S \rangle$ with any filter, top-hat included. Let consider the correlation of the $R$--derivative of density contrast field filtered at different smoothing lengths
\begin{equation}
\begin{split}
\left\langle \frac{\partial \delta(R_1)}{\partial R_1} \frac{\partial \delta(R_2)}{\partial R_2} \right\rangle &= \frac{\partial }{\partial R_1} \frac{\partial }{\partial R_2} \left\langle \delta(R_1) \delta(R_2) \right\rangle \\
&= \frac{\partial }{\partial R_1} \frac{\partial }{\partial R_2} C(R_1,R_2)\,,
\end{split}
\end{equation}
where $C(R_1,R_2)$ is the covariance defined in Eq.~\eqref{eq:C_ij_def}, which is a well defined quantity. From that, the variance of the scale derivative of the density contrast field can be easily derived as
\begin{equation}\label{eq:der_variance}
\left\langle \left( \frac{\partial \delta(R)}{\partial R} \right)^2 \right\rangle = \left. \frac{\partial }{\partial R_1} \frac{\partial }{\partial R_2} C(R_1,R_2) \right\vert_{R_1=R_2=R}.
\end{equation}
This relation can extended at any order in the derivative of the filtered field,
\begin{equation}
\left\langle \frac{\partial^n \delta(R)}{\partial R^n} \frac{\partial^m \delta(R)}{\partial R^m} \right\rangle = \left. \frac{\partial^n }{\partial R_1^n} \frac{\partial^m }{\partial R_2^m} C(R_1,R_2) \right\vert_{R_1=R_2=R}.
\end{equation}
To further confirm the above relation, we measure the ensemble variance over many MC realisations of the field $\partial \delta(R) / \partial R$, obtained by finite differentiation of the field $\delta(R)$ from Eq.~\eqref{eq:delta_sigle_cholesky} with the top-hat filter. Fig.~\ref{fig:variance_derivative} shows the results: $\langle (\partial \delta(R) / \partial R)^2 \rangle$ computed as the ensemble variance of MC realisations (orange line); same quantity computed with Eq.~\eqref{eq:der_variance} (black dashed line); $\sigma^2(R)=\langle \delta^2(R) \rangle$ (blue line).

\section{Derivation of the analytical multiplicity function}\label{ap:properties}

The joint distribution of $\delta_S$, $\delta'$, and $\delta_s$ is a multivariate normal distribution
\begin{equation}\label{eq:p_joint_def}
p(\delta_S,\delta'_S,\delta_s ) = \frac{1}{(2 \pi)^{3/2} \sqrt{\det {\bf \Sigma}}} \exp \left\{-\frac{1}{2} (\delta_S,\delta'_S,\delta_s ) \begin{array}{c}
\text{\LARGE\ensuremath {\bf \Sigma}}^{-1} 
\end{array}\left(
\begin{array}{c}
\delta_S\\
\delta'_S\\
\delta_s 
\end{array}
\right)
\right\}
\end{equation}
where ${\bf \Sigma}$ is the covariance matrix of Eq.~\eqref{eq:covariance}. The exponent of Eq.~\eqref{eq:p_joint_def} reads
\begin{equation}
\frac{-1}{2 \det {\bf \Sigma}} \left[ \Gamma_{BB} B_S^2
 + \Gamma_{\delta' \delta'} \delta'^2_S + \Gamma_{\delta \delta} \delta^2_s    + 2\Gamma_{B \delta'} B_S \delta'_S + 2\Gamma_{B \delta} B_S \delta_s + 2\Gamma_{\delta \delta'} \delta'_S \delta_s \right] 
\end{equation}    
where
\begin{equation}
\det {\bf \Sigma} = sSD - SC'^2 + C C' - s/4 - D C^2 
\end{equation}
and $\Sigma_{ij}^{-1} = \Gamma_{ij} / \det {\bf \Sigma}$, with $\Gamma_{ij}$ listed in Eq.~\eqref{eq:gammas}. The integral with respect $\delta'$ of Eq.~\eqref{eq:def_f_new} is
\begin{align}
\int_{B'}^\infty & {\rm d} \delta' p(B_S,\delta'_S,\delta_s)  (\delta'_S - B'_S) =  \\
&\frac{1}{(2 \pi)^{3/2} \sqrt{\det {\bf \Sigma}}} \left(\vbox to 25pt{}\right. \frac{\det {\bf \Sigma}}{\Gamma_{\delta' \delta'}} \exp \left[ -\frac{\Gamma_{\delta' \delta'} B'^2_S + 2 \kappa B'_S}{2 \det {\bf \Sigma}} \right] +\\
& \sqrt{\frac{\pi}{2} \frac{\det {\bf \Sigma}}{\Gamma_{\delta' \delta'}}} \left( B'_S + \frac{\kappa}{\Gamma_{\delta' \delta'}} \right)\exp \left[ \frac{\kappa^2/\Gamma_{\delta' \delta'}}{2 \det {\bf \Sigma}} \right] \left\{ {\rm erf} \left[ \sqrt{\frac{\Gamma_{\delta' \delta'}}{2 \det {\bf \Sigma}}} \left( B'_S + \frac{\kappa}{\Gamma_{\delta' \delta'}} \right) \right] - 1 \right\} \left.\vbox to 25pt{}\right)\,,  \nonumber
\end{align}
where $\kappa=\Gamma_{B \delta'} B_S + \Gamma_{\delta' \delta} \delta_s$.
The integration of the above expression with respect to $\delta$ can be reduced into sums and products of integrals of the form 
\begin{enumerate}
    \item $\int {\rm d} x \, e^{-\alpha(x-\beta)^2}$
    \item $\int {\rm d} x \, x e^{-\alpha(x-\beta)^2}$
    \item $\int {\rm d} x \, e^{-\alpha(x-\beta)^2} {\rm erf} \big[\gamma (x + \epsilon) \big]$
    \item $\int {\rm d} x \, x e^{-\alpha(x-\beta)^2} {\rm erf} \big[\gamma (x + \epsilon) \big]$
\end{enumerate}
which all have an explicit solution as combinations of Gaussians and error functions, except for the last case. Taking into account the following relations
\begin{itemize}
\item $\Gamma_{\delta \delta} \Gamma_{BB} - \Gamma_{B \delta}^2 = D \det {\bf \Sigma}$

\item $\Gamma_{\delta' \delta'} \Gamma_{\delta \delta} - \Gamma_{\delta' \delta}^2 = S \det {\bf \Sigma}$

\item $\Gamma_{\delta' \delta'} \Gamma_{BB} - \Gamma_{B \delta'}^2 = s \det {\bf \Sigma}$

\item $\Gamma_{B \delta'} \Gamma_{\delta \delta} - \Gamma_{B \delta} \Gamma_{\delta' \delta} = -\frac{1}{2} \det {\bf \Sigma}$

\item $\Gamma_{B \delta} \Gamma_{\delta' \delta'} -  \Gamma_{\delta' \delta} \Gamma_{B \delta'} = -C \det {\bf \Sigma}$
\end{itemize}
we obtain Eq.~\eqref{eq:Ps_final}.

\section{Global and boundary behaviour}\label{ap:limit}

\subsection{$ s \rightarrow S$ limit}
Eq.~\eqref{eq:Ps_final} behaves as $0/0$ as $s \rightarrow S$. The quantities $\Gamma_{ij}$ and $\det {\bf \Sigma}$ can be written in terms of $s$, $S$, $D$, $C$, $C'$. In particular, $C$ and $C'$ can be expanded around $S$ as
\begin{equation}
\begin{split}
C(s) &= \sum_n  \frac{(s - S)^n}{n!} \langle \delta \delta^{(n)} \rangle|_{s=S} \\
C'(s) &= \sum_n  \frac{(s - S)^n}{n!} \langle \delta' \delta^{(n)} \rangle|_{s=S},
\end{split}
\end{equation}
where $(n)$ denote the $n^{\rm th}-$derivative with respect to $s$.
We consider the correlation of the fields, both eventuated at the same $s=S$, up to the $4^{\rm th}-$derivative of the second field. 
\begin{itemize}
\item $ \langle \delta \delta' \rangle = \frac{1}{2} \frac{\partial}{\partial S} \langle \delta^2 \rangle = \frac{1}{2} \frac{\partial}{\partial S} S = \frac{1}{2}$

\item $ \langle \delta \delta'' \rangle$: $\frac{\partial}{\partial S} \langle \delta \delta' \rangle = 0 = \langle \delta \delta'' \rangle + \langle \delta'^2 \rangle \Rightarrow \langle \delta \delta'' \rangle =\langle \delta'^2 \rangle= -D$

\item $ \langle \delta' \delta'' \rangle = \frac{1}{2} \frac{\partial}{\partial S} \langle \delta'^2 \rangle = \frac{1}{2} \frac{\partial}{\partial S} D = \frac{1}{2} D'$

\item $ \langle \delta \delta''' \rangle$: $\frac{\partial^2}{\partial S^2} \langle \delta \delta' \rangle = 0 = \frac{\partial}{\partial S} \left[ \langle \delta \delta'' \rangle + \langle \delta'^2 \rangle \right] =\langle \delta \delta''' \rangle + 3 \langle \delta' \delta'' \rangle\Rightarrow \langle \delta \delta''' \rangle = -\frac{3}{2} D'$

\item $ \langle \delta' \delta''' \rangle$: $\frac{\partial}{\partial S} \langle \delta' \delta''' \rangle =  \langle \delta' \delta''' \rangle + \langle \delta''^2 \rangle  \Rightarrow \langle \delta' \delta''' \rangle = \frac{1}{2} D'' - \langle \delta''^2 \rangle $

\item $ \langle \delta \delta'''' \rangle$: $ \frac{\partial}{\partial S} \left[ \langle \delta \delta''' \rangle + 3 \langle \delta' \delta'' \rangle \right] = 0= \langle \delta \delta'''' \rangle + 4\langle \delta' \delta''' \rangle + 3\langle \delta''^2 \rangle\Rightarrow \langle \delta \delta'''' \rangle = \langle \delta''^2 \rangle + 2 D''$

\item $ \langle \delta' \delta'''' \rangle$: $\frac{\partial^3}{\partial S^3} \langle \delta'^2 \rangle =  6\langle \delta'' \delta''' \rangle + 2\langle \delta' \delta'''' \rangle  \Rightarrow \langle \delta' \delta'''' \rangle = \frac{1}{2} D''' - \frac{3}{2} \frac{\partial}{\partial S} \langle \delta''^2 \rangle$

\end{itemize}
where $\langle \delta''^2 \rangle$ can be evaluated following Appendix~\ref{ap:compute_W}. 
Using these relations, it is possible to Fourier expand $\Gamma_{ij}$, Eq.~\eqref{eq:gammas}, and ${\det {\bf \Sigma}}$, Eq.~\eqref{eq:covariance}, in the limit $s\rightarrow S$. The first non-zero coefficient of the Taylor expansion around $S$ of ${\det {\bf \Sigma}}$ is the $4^{\rm th}$ one, i.e. ${\det {\bf \Sigma}} \propto (s-S)^4$ as $s \rightarrow S$. 

For convenience, we rewrite Eq.~\eqref{eq:Ps_final}
as 
\begin{equation}\label{eq:Ps_Ii}
{\cal P}(s) = I_1 + I_2 + e^{-B^2_S/2S} \left( I_3 + I_4 \right),
\end{equation}
where the 4 terms $I_i$ are in the same order of the ones appearing in Eq.~\eqref{eq:Ps_final}.
$I_1$, the first term on the right-hand-side of Eq.~\eqref{eq:Ps_final}, i.e. the second and third lines, goes to zeros, as the exponential and $\Gamma_{\delta \delta}$ are constant, while $\Gamma_{\delta' \delta'} \propto (s-S)^2$. Therefore, the quantity ${\det {\bf \Sigma}} / \Gamma_{\delta' \delta'} \propto (s-S)^2$, which multiplies a finite quantity, i.e. $I_1 \rightarrow 0$  as $s \rightarrow S$. 
Expanding up to the second order, the second term, $I_2$, i.e. the fourth line of Eq.~\eqref{eq:Ps_final}, becomes
\begin{equation}
I_2  \rightarrow \frac{\sqrt{\Gamma_{\delta \delta}}}{4 \pi S} \exp \left( - \frac{D B_S^2 + SB' - B_SB}{2 \Gamma_{\delta \delta}} \right). 
\end{equation}
The third and fourth terms are multiplied by $e^{-B^2_S/2 S}$, which is finite and constant.
The argument of the error function in the second term inside the curly brackets, $I_4$, is
\begin{equation}
\sqrt{\frac{S}{2 \Gamma_{\delta' \delta'}}} \left(B(s) - \frac{C}{S}B_S \right)  \rightarrow \sqrt{\frac{S}{2 \Gamma_{\delta \delta}}}\left(\frac{B_S}{2S} - B' \right).
\end{equation}
Finally, the last quantity is
\begin{equation}\label{eq:limit_I3}
I_3 = \frac{\det {\bf \Sigma}}{2 \pi \sqrt{\Gamma_{\delta' \delta'}}\Gamma_{\delta' \delta}} {\cal F}(x,\alpha,\beta)
\end{equation}
where ${\det {\bf \Sigma}} / (\sqrt{\Gamma_{\delta' \delta'}}\Gamma_{\delta' \delta}) \propto (s-S)^2$, and ${\cal F}(x,\alpha,\beta)$ is defined in Eq.~\eqref{eq:Fxab}. It can be noticed from Eq.~\eqref{eq:x_alpha_beta} that
\begin{align}
x \rightarrow \sqrt{\frac{\Gamma_{\delta \delta}}{2 \det {\bf \Sigma}_{(4)}}} &\left[ \left(\frac{D'}{2} - 
D^2 \right) \frac{B_s}{2 \Gamma_{\delta \delta}} \right. -  \\
&\left.(S D' + D) \frac{B'}{4 \Gamma_{\delta \delta}} + \frac{B''}{2}\right], \nonumber
\end{align}
where $\det {\bf \Sigma}_{(4)}$ is the $4^{\rm th}$-order coefficient of the Taylor expansion. The above equation is a finite quantity, while $\alpha \propto (s-S)^2$ and $\beta \propto (s-S)^{-1}$. It follows that in the limit $s \rightarrow S$, the exponential term appearing in the integral of Eq.~\eqref{eq:Fxab} behaves as a Gaussian infinitely wide and centered at $-\infty$. Note, however, that the width of this Gaussian, i.e. $\sqrt{2/\alpha}$, and the center position, i.e. $-\beta$, compensate for each other,
\begin{equation}
\alpha \beta^2 = S \frac{\Gamma_{\delta' \delta'}}{\Gamma_{\delta' \delta}^2} \left( \frac{B_S}{2S} - B'\right)^2 \rightarrow \frac{S}{\Gamma_{\delta \delta}} \left( \frac{B_S}{2S} - B'\right)^2.
\end{equation}
This means that the exponential contributes to the integral as a finite positive quantity in the entire range from $-\infty$ to $x$. As a consequence, the upper integration limit, $x$, is negligible and can be considered $x \rightarrow 0$. The same argument applies to the error function: the scale range in which it differs from -1 is negligible, therefore we can consider ${\rm erf}(t) \rightarrow -1$ over all the scales of interest. In this limit, the integral of Eq.~\eqref{eq:Fxab} becomes
\begin{align}
{\cal F}(x,\alpha,\beta) &\rightarrow -\int_{-\infty}^0 {\rm d}t \, t \, e^{-\alpha(t+\beta)^2} \\ 
&=\frac{e^{-\alpha \beta^2}}{2 \alpha} + \frac{\sqrt{\pi}}{2}\frac{\beta}{\sqrt{\alpha}} \left[ 1 + {\rm erf} \left( \sqrt{\alpha} \beta\right)\right]\,, \nonumber
\end{align}
from which, the limit for $s \rightarrow S$ of Eq.~\eqref{eq:limit_I3}, is
\begin{equation}
I_3 \rightarrow \frac{\sqrt{\Gamma_{\delta \delta}}}{4 \pi S}e^{-\alpha \beta^2} + \frac{{\rm erf}(\sqrt{\alpha}\beta) + 1}{4 \sqrt{s \pi S}}\left( \frac{B_S}{2S} - B'\right)\,.
\end{equation}
Substituting the four $I_i$ terms computed here in Eq.~\eqref{eq:Ps_Ii}, we find that Eq.~\eqref{eq:Ps_final}, in the limit $s \rightarrow S$, becomes
\begin{align}\label{eq:limit_S_final}
\lim_{s\rightarrow S} {\cal P}(s) = e^{-B^2_S / 2S}  &\left\llbracket \frac{\sqrt{\Gamma_{\delta \delta}}}{2 \pi S} \exp \left[ -\frac{S}{2\Gamma_{\delta \delta}} \left(\frac{B_S}{2S} - B'_S \right)^2\right]\right. +  \\
&\quad\left. \frac{\left(B_S/2S - B'_S \right)}{2\sqrt{2 \pi S}}  \left\{{\rm erf}\left[\sqrt{\frac{S}{2\Gamma_{\delta \delta}}} \left(\frac{B_S}{2S} - B'_S \right)\right]+1\right\} \right\rrbracket \,, \nonumber 
\end{align}
where we use the fact that the exponent appearing in $I_1$ can be factorized as
\begin{equation}
\frac{D B_S^2 + SB' - B_SB}{2 \Gamma_{\delta \delta}} = \frac{S}{2\Gamma_{\delta \delta}} \left(\frac{B_S}{2S} - B'_S \right)^2 +\frac{B_S}{2 S} .
\end{equation}

\subsection{$s \rightarrow 0$ limit}
We now explore the $s \rightarrow 0$ limit of ${\cal P}(s)$, Eq.~\eqref{eq:Ps_final}. Note that $S > s$ by construction, so $R_S < R_s$. Considering $S$ as a finite quantity,  $R_S \ll R_s\rightarrow \infty$ as $s \rightarrow 0$, entailing that
\begin{align}
C &= \int \frac{{\rm d} k}{2 \pi^2}  k^2 P(k) W(kR_S) W^*(kR_s) \\
&\rightarrow \int \frac{{\rm d} k}{2 \pi^2}  k^2 P(k) W(kR_s) = {\cal O}(s) \,,
\end{align}
since $W(kR_S) \sim 1$ for $k\ll 1/R_S$, while $W(kR_s) \rightarrow 0$ as $k \gtrsim 1/R_s$ with $1/R_s \ll 1/R_S$. 
Due to the phase shift in the derivative of the filter function $W(kR)$ with respect to $R$, $\partial W(kR_S) / \partial R_S \rightarrow 0$ for $k\ll 1/R_S$; while $|W(kR_s)| > 0$
for $k$ on the range between $0$ and a few times $\sim 1/R_s$, and $1/R_s \ll 1 / R_S$. It follows that $C' \rightarrow 0$ at higher order than $s$ and $C$. Taylor expanding up to the first order in $s$ around $s=0$, the following quantities becomes
\begin{equation}\label{eq:gammas_at_0}
\begin{split}
&\det {\bf \Sigma} \simeq s (S D -1/4) = s \Gamma_{\delta \delta} \\
&\begin{array}{lcl}
\Gamma_{BB} \simeq sD & & \Gamma_{B\delta'} \simeq - s/2 \\
\Gamma_{\delta' \delta'} \simeq sS &  & \Gamma_{B\delta} \simeq - C D \\
\Gamma_{\delta \delta} = SD - 1/4 &  & \Gamma_{\delta' \delta} \simeq C/2  \,,
\end{array}
\end{split}
\end{equation}
where $\Gamma_{\delta \delta}$ is a constant quantity.

We start considering the first term on the right-hand side of Eq.~\eqref{eq:Ps_final}, $I_1$, i.e. the second and third lines. The argument of the error function goes as
\begin{align}
&\sqrt{\frac{\Gamma_{\delta \delta}}{2 \det {\bf \Sigma}}} \left(B(s) + \frac{\Gamma_{B \delta}}{\Gamma_{\delta \delta} }B_S + \frac{\Gamma_{\delta' \delta}}{\Gamma_{\delta\delta} }B'_S \right) \nonumber \\
&\rightarrow \frac{1}{\sqrt{2 s}} \left( B(s) - \frac{D B_S - B'_S/2}{\Gamma_{\delta \delta}} C\right).
\end{align}
The behaviour at $s \rightarrow 0$ depends on the barrier shape. Taking into account a barrier of the form Eq.~\eqref{eq:moving_b}, in both the cases $B(s) \rightarrow {\rm const}$, i.e. $\gamma \geq 0$, and $B(s) \rightarrow \infty$, i.e. $\gamma < 0$, as $s \rightarrow 0$, the above quantity goes to $\infty$, and the corresponding error function to 1. Using Eqs.~\eqref{eq:gammas_at_0}, the factor multiplying the exponential and the curly brackets behaves as
\begin{equation}
\frac{\det {\bf \Sigma}}{4 \pi \sqrt{\Gamma_{\delta \delta}}\Gamma_{\delta' \delta'}} \rightarrow \frac{\sqrt{\Gamma_{\delta \delta}}}{4 \pi S},
\end{equation}
and, therefore, since the exponential term does not depend on $s$, the first term of Eq.~\eqref{eq:Ps_final} at $s \rightarrow 0$ is
\begin{equation}
I_1 \rightarrow \frac{\sqrt{\Gamma_{\delta \delta}}}{2 \pi S}\exp \left[ -\frac{SB'^2_S + D B_S^2 - B'_S B_S}{2 \Gamma_{\delta \delta}} \right]
\end{equation}

Let us consider the second term, $I_2$, i.e. the fourth line of Eq.~\eqref{eq:Ps_final}. The factor multiplying the exponential behaves as
\begin{equation}
\frac{\Gamma_{\delta' \delta}}{4 \pi S \sqrt{\Gamma_{\delta' \delta'}}} \rightarrow \frac{C}{8 \pi S \sqrt{S s}} \propto \sqrt{s},
\end{equation}
where the dependence $\propto \sqrt{s}$ is obtained Taylor expanding $C$ up to the first order in $s$, $C \simeq \partial_s C|_{s=0} s$. The 3 terms in the exponential behave as
\begin{align}
S B^2(s) / (2 \Gamma_{\delta' \delta'}) &\rightarrow B^2(s)/s \rightarrow \infty \nonumber \\ 
s B_S^2 / (2 \Gamma_{\delta' \delta'}) &\rightarrow B_S^2 / 2S = {\rm const} \\
C B_S B(s) / \Gamma_{\delta' \delta'} &\rightarrow \partial_s C|_{s=0} B_S B(s) / S\rightarrow {\rm const \ or\ } \infty, \nonumber
\end{align}
where the first equation goes always to $\infty$ for any behaviour of $B(s)$, considering a form as Eq.~\eqref{eq:moving_b}, while the last equations may got either to a constant value or $\infty$ depending on $B(s)$ behaviour. In any case, the exponent term tends to $-\infty$ as $s \rightarrow 0$, so $I_2 \rightarrow 0$.

The last term on the right-hand side of Eq.~\eqref{eq:Ps_final}, i.e. $4^{\rm th}-5^{\rm th}$ lines of Eq.~\eqref{eq:Ps_final}, is composed of a constant exponential term multiplying the sum $I_3 + I_4$. The argument of the error function of $I_4$, i.e. $5^{\rm th}$ line of Eq.~\eqref{eq:Ps_final}, in the limit $s \rightarrow 0$, behaves as
\begin{equation}
\sqrt{\frac{S}{2 \Gamma_{\delta' \delta'}}} \left(B(s) - \frac{C}{S}B_S \right) \rightarrow \frac{B(s) - C B_S / S} {\sqrt{2 s}} \propto \frac{B(s)}{\sqrt{s}} ,
\end{equation}
which goes to $\infty$ for any behaviour of $B(s)$, if the functional form of Eq.~\eqref{eq:moving_b} is considered. It follows that the corresponding error function goes to 1 and
\begin{equation}
I_4 \rightarrow \frac{B_S/2S - B'_S}{2 \sqrt{2 \pi S}}
\end{equation}
as $s \rightarrow 0$.
The remaining term in the curly brackets is $I_3$, which is the product of
\begin{equation}
\frac{2 \det \Sigma}{4 \pi \sqrt{\Gamma_{\delta' \delta'}} \Gamma_{\delta' \delta}} \rightarrow \frac{\Gamma_{\delta \delta}}{C}\sqrt{\frac{s}{S}} \rightarrow \frac{\Gamma_{\delta \delta}}{\sqrt{S} \partial_sC|_{s=0}}\frac{1}{\sqrt{s}},
\end{equation}
multiplying ${\cal F}(x,\alpha,\beta)$. Let us consider the $x$, $\alpha$, $\beta$ quantities of Eqs.~\eqref{eq:x_alpha_beta}. The $x$ quantity behaves as
\begin{equation}
x \rightarrow \frac{\partial_s C|_{s=0}}{2 \sqrt{2 S \Gamma_{\delta \delta}}} \left( B(s) - \frac{B_S}{\partial_s C} + \frac{2 S}{\partial_s C} B'_S\right),
\end{equation}
which goes to $\infty$ if $B(s) \rightarrow \infty$ as $s \rightarrow 0$, or to a constant value otherwise. The other quantities behave as
\begin{equation}
\alpha = \frac{S \det {\bf \Sigma}}{\Gamma_{\delta' \delta}^2} \rightarrow \frac{4 \Gamma_{\delta \delta} S s}{C^2} \rightarrow \frac{4 \Gamma_{\delta \delta} S}{(\partial_s C|_{s=0})^2} \frac{1}{s}
\end{equation}
and
\begin{equation}
\beta \rightarrow \sqrt{\frac{S}{2 \Gamma_{\delta \delta}}} \left( \frac{B_S}{2 S} - B'_s \right) = {\rm const}
\end{equation}
as $s \rightarrow 0$. 
From the above quantities, it follows that the exponential $e^{-\alpha(t+\beta)^2}$ appearing in the integrand of ${\cal F}(x,\alpha,\beta)$, Eq.~\eqref{eq:Fxab}, behaves as an infinitesimal narrow Gaussian centered in $-\beta$. In the regime we are considering, it can be shown comparing $x$ and $\beta$, together with Eqs.~\eqref{eq:gammas_at_0}, that $-\beta < x$, and therefore
\begin{align}\label{eq:limit_0_Fxab}
{\cal F}(x,\alpha,\beta) &\rightarrow \left. t \, {\rm erf} \right|_{t=-\beta} \int_{-\infty}^\infty {\rm d} t e^{-\alpha t^2} = \beta {\rm erf} (\beta) \sqrt{\frac{\pi}{\alpha}} \nonumber \\
& \rightarrow \frac{B_S/2S - B'_S}{2 \sqrt{2 \pi S}} {\rm erf} \left[ \sqrt{\frac{S}{2 \Gamma_{\delta \delta}}} \left( \frac{B_S}{2 S} - B'_s \right) \right] .
\end{align}
Now it is possible to sum all the terms of Eq.~\eqref{eq:Ps_final} in the $s \rightarrow 0$ limits, from which it follows that $\lim_{s\rightarrow 0} {\cal P}(s) = \lim_{s\rightarrow S} {\cal P}(s)$, as appearing in Eq.~\eqref{eq:limit_S_final}.

\begin{figure}[t!]
\centering
\includegraphics[width=0.67\textwidth]{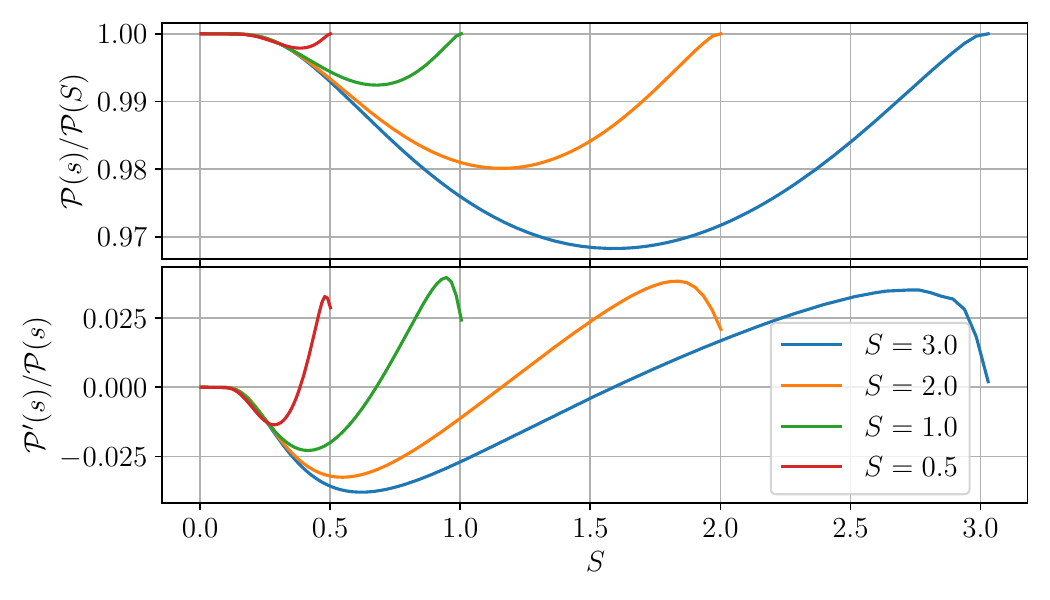}
\caption{Upper panel: ${\cal P}(s) / {\cal P}(S)$ for various $S$, listed in the legend. Bottom panel: ${\cal P}'(s) / {\cal P}(s)$, the colour refers to the same $S$ as in the upper panel.}
\label{fig:PS_beharior}
\end{figure}

\subsection{Derivative of $ {\cal P}(s)$ and global behaviour}
We now consider the global behaviour of ${\cal P}(s)$.
The upper panel of Fig.~\ref{fig:PS_beharior}, shows the behaviour of ${\cal P}(s)$ with respect to its value at the crossing scale, ${\cal P}(S)$, for various crossing scales $S$ corresponding to different colour. The ${\cal P}(s)$ function shown here is the same as that described in Sec.~\ref{sec:multipl_func}, at $z=0$. It can be seen that $\lim_{s\rightarrow 0} {\cal P}(s) = \lim_{s\rightarrow S} {\cal P}(s)$, as previously derived, and that ${\cal P}(s)$ slowly varies in the range $0-S$, from sub- to a few percent level, depending on the corresponding crossing scale $S$. The lower panel shows the derivative ${\cal P}'(s)$ with respect to ${\cal P}(s)$, for various crossing scales $S$. The derivative value is always within a few percent level the corresponding ${\cal P}(s)$ function. This further confirms that ${\cal P}(s)$ is a slowly varying function. From these properties, it follows that the integral of ${\cal P}(s)$ over $s$ can be approximated as $\int_0^S {\cal P}(s) {\rm d}s \simeq S\lim_{s\rightarrow 0} {\cal P}(s) = S\lim_{s\rightarrow S} {\cal P}(s)$ in the limit of small $S$, obtaining Eq.~\eqref{eq:multipl_integr_approx}.

\section{Local bias expansion}\label{ap:local_bias}

For any multiplicity function with dependence on the formation threshold, the corresponding local bias expansion naturally arises in the peak-background split framework, as a response of the formation threshold on the background variation~\citep{kaiser_1984,bardeen_1986,cole_kaiser_1989,mo_white_1996,mo_jing_white_1997,sheth_tormen_1999}. This is the case for the effective barrier approach explored in this paper. 
A modification in the background density due to a long-wavelength perturbation can be treated as a modification of the formation threshold,
$f(S,\delta|\Delta_l) = f(S,\delta-\Delta|0)$, where $\Delta_l$ is the long-wavelength perturbation. Therefore, the local bias expansion can be written as
\begin{equation}
b_{\rm N}(S) = \frac{(-1)^N}{f(S)} \frac{\partial^N f(S)
}{\partial \delta^N} .
\end{equation}
Since the dependence with respect to the formation threshold is contained in the moving barrier $B_S$, we separate the derivative of $B_S$ with respect to $\delta$ and the derivative of $f(S)$ with respect to the barrier. From the above equations, the first bias terms are
\begin{align}
b_1(S) &= \frac{-1}{f(S)}\frac{\partial B_S}{\partial \delta}\frac{\partial}{\partial B_S} f(S) \, ,  \\
b_2(S) &= \frac{1}{f(S)} \left[ \frac{\partial^2 B_S}{\partial \delta^2} \frac{\partial}{\partial B_S} + \left(\frac{\partial B_S}{\partial \delta}\right)^2 \frac{\partial^2}{\partial B_S^2
}  \right] f(S) \,, \nonumber \\
b_3(S) &= \frac{-1}{f(S)} \Bigg[ \frac{\partial^3 B_S}{\partial \delta^3} \frac{\partial}{\partial B_S}  + 3 \frac{\partial B_S}{\partial \delta} \frac{\partial^2 B_S}{\partial \delta^2} \frac{\partial^2}{\partial B_S^2} +  \left(\frac{\partial B_S}{\partial \delta}\right)^3 \frac{\partial^3}{\partial B_S^3} \Bigg] f(S) \,. \nonumber
\end{align}
Note that the barrier is formally a function of two variables, $B_S=B(S,\delta)$. When implementing the above equations with the analytical multiplicity functions presented in this work, Eqs.~\eqref{eq:Ps_final}-\eqref{eq:multipl_integr_approx}, this has to be taken into account when computing the derivative of $B'$ with respect to the barrier, which reads
\begin{align}
\frac{\partial B'_S}{\partial B_S} &= \left. \frac{\partial B'_S}{\partial B_S} \right\vert_S = \left[\frac{\partial B_S}{\partial \delta}\right]^{-1} \frac{\partial B'_S}{\partial \delta} , \\
\frac{\partial^2 B'_S}{\partial B_S^2} &= \left[\frac{\partial B_S}{\partial \delta}\right]^{-1} \frac{\partial^2 B'_S}{\partial \delta^2} - \left[\frac{\partial B_S}{\partial \delta}\right]^{-2} \frac{\partial^2 B_S}{\partial \delta^2} \frac{\partial B'_S}{\partial \delta} , \nonumber \\
\frac{\partial^3 B'_S}{\partial B_S^3} &= \left[\frac{\partial B_S}{\partial \delta}\right]^{-1} \frac{\partial^3 B'_S}{\partial \delta^3} -2\left[\frac{\partial B_S}{\partial \delta}\right]^{-2} \frac{\partial^2 B_S}{\partial \delta^2} \frac{\partial^2 B'_S}{\partial \delta^2} + \nonumber\\
& \qquad \left\{  2\left[\frac{\partial B_S}{\partial \delta}\right]^{-3} \left[\frac{\partial^2 B_S}{\partial \delta^2}\right]^2  -\left[\frac{\partial B_S}{\partial \delta}\right]^{-2} \frac{\partial^3 B_S}{\partial \delta^3} \right\} \frac{\partial B'_S}{\partial \delta} \nonumber \,.
\end{align}
When considering the approximated multiplicity function, Eq.~\eqref{eq:multipl_integr_approx}, we obtain
\begin{align}
\frac{\partial f(S)}{\partial B_S} &= -\frac{B_S}{S}f(S) + \frac{e^{-B_S^2/2S}}{2\sqrt{2 \pi S}} \partial_B \Delta_B \times \left[ {\rm erf} \left(\sqrt{\frac{S}{2\Gamma_{\delta \delta}}}\Delta_B \right) +1 \right] ,\nonumber\\
\frac{\partial^2 f(S)}{\partial B_S^2} &=  \left(\frac{B_S^2}{S^2} - \frac{1}{S}\right)f(S) +  \frac{e^{-B_S^2/2S}}{\sqrt{2 \pi S}} \left\{ (\partial_B \Delta_B)^2 \sqrt{\frac{S}{2 \pi \Gamma_{\delta \delta}}} \exp\left(-\frac{S\Delta_B^2}{2 \Gamma_{\delta\delta}} \right) + \right.  \nonumber\\
& \left.\frac{1}{2}\left(\partial_B^2 \Delta_B - 2\frac{B}{S} \partial_B\Delta_B  \right) \left[ {\rm erf} \left(\sqrt{\frac{S}{2 \Gamma_{\delta\delta}}}\Delta_B\right) +1 \right] \right\} ,\\
\frac{\partial^3 f(S)}{\partial B_S^3} &=  \left( \frac{3B}{S^2} - \frac{B^3}{S^3} \right) f(S) +  \frac{e^{-B_S^2/2S}}{\sqrt{2 \pi S}} \times \nonumber\\
&\Bigg\{ \left[ 3 \partial^2_B\Delta_B \partial_B\Delta_B -\frac{3B}{S} (\partial_B \Delta_B)^2 - \frac{S}{\Gamma_{\delta \delta}} \Delta_B(\partial_B\Delta_B)^3\right]  \sqrt{\frac{S}{2 \pi \Gamma_{\delta \delta}}} \exp\left(-\frac{S\Delta_B^2}{2 \Gamma_{\delta\delta}} \right) + \nonumber\\
&\quad\frac{1}{2}\left[ \left( \frac{3B}{S^2} - \frac{3}{S} \right) \partial_B \Delta_B -\frac{3B}{S} \partial^2_B \Delta_B + \partial^3_B \Delta_B \right] \left[ {\rm erf} \left(\sqrt{\frac{S}{2 \Gamma_{\delta\delta}}}\Delta_B\right) +1 \right] \Bigg\} \nonumber
\end{align}
where $\Delta_B = (B_S/2S - B'_S)$, $ \partial_B \Delta_B = 1/2S - \partial B'_S / \partial B_S$, $ \partial^2_B \Delta^2_B = - \partial^2 B'_S / \partial B_S^2$, and $ \partial^3_B \Delta^3_B = - \partial^3 B'_S / \partial B_3^2$.

\section{Measuring the moving barrier}\label{ap:mcmc_barrier}

In this Section we describe how the MCMC to fit $\alpha$, $\beta$, and $\gamma$ described in Sec.~\ref{sec:effective_barrier} is performed. We proceed in various steps.
First, we obtain the multiplicity functions of the VSF measured in both the DEMNUni ICs and HR-DEMNUni ICs, for each of the void formation thresholds considered, Tab.~\ref{tab:thresholds}. Then we consider the minimum useful void radius, in order to avoid scales in which voids are noise dominated. This minimum radius depends simultaneously on the spatial resolution of the simulation ICs, the void finder grid size, and the threshold value. In particular, we found that the void radius cannot be smaller than $\sim 5$ times the voxel side of the void finder grid. We found that 1000 grid division per box side guarantees the best possible spatial resolution given the number of particles of DEMNUni simulations, for both the standard and HR ICs, see Sec.~\ref{sec:effective_barrier}. This results in a minimum useful radius of $\sim 10~h^{-1}{\rm Mpc}$ for DEMNUni ICs and $\sim 2.5~h^{-1}{\rm Mpc}$ for HR-DEMNUni ICs. Beyond the effect of the grid, the minimum useful void radius can be evaluated as a function of the scale corresponding to the Poissonian noise associated to the number of particles in a void with a given underdensity threshold. 
The expected number of particles in a sphere with radius $R$ is $N(R) = 4 \pi n_{\rm p} R^3/3$, where $n_{\rm p}$ is the mean number density of particles in the simulation box. The associated Poissonian uncertainty is $\sqrt{N}$. The number of particles contained in a void with formation threshold $\delta(z_{\rm IC})$ and radius $R$ detected by the void finder is $\Delta N(R) = \delta(z_{\rm IC}) \, 4 \pi n_{\rm p} R^3/3$. We can define the radius $R_{\rm P}(P)$ as the radius of the void containing a number of particles equal to $P$ times the Poissonian uncertainty associated with the sphere with radius $R_{\rm P}(P)$: $\Delta N[R_{\rm P}(P)] = P \sqrt{N[R_{\rm P}(P)]}$. We use this radius as an estimator of the Poissonian uncertainty associated to the detection of a void with threshold $\delta(z_{\rm IC})$ and radius $R_{\rm P}(P)$
\begin{equation}
R_{\rm P}(P) = \left( \frac{P}{\delta(z_{\rm IC})} \right)^{2/3} \left( \frac{3}{4 \pi n_{\rm p}} \right)^{1/3},
\end{equation}
Exploring various minimum radii in our analysis, we find that the corresponding posterior distributions of the MCMC are stable for radius greater than $R_{\rm P}(\sim 2)$, i.e. containing a number of particles that is at least 2 times the Poissonian uncertainty of the mean number density. Tab.~\ref{tab:Rmin} lists the minimum radius used for each threshold explored.

To speed up the analysis, we start exploring the parameter space of the effective barrier of Eq.~\eqref{eq:moving_b} performing a MCMC with the approximated analytical multiplicity function, Eq.~\eqref{eq:multipl_integr_approx}. We consider a Gaussian likelihood function, 
\begin{equation}\label{eq:likelihood}
\log \left[ {\cal L}({\cal D} | \Theta) \right] = -\frac{1}{2} \sum_{i} \left[\left(n^{\cal D}_i - n^T_i(\Theta) \right) ^2 \Sigma^{-1}_i + \log ( \Sigma_i ) \right],
\end{equation}
where the theoretical model is the theoretical multiplicity function integrated in the radius bins considered $n^T_i(\Theta) = \int_{S(R_i)}^{S(R_{i-1})}f(S) {\rm d} S$ with $\Theta = \{\alpha,\beta,\gamma\}$; the data vector with elements $n^{\cal D}_i$, which is the measured multiplicity function in the various $i^{\rm th}$ bin, $n^{\cal D}_i = N_i  4 \pi (\langle R \rangle_i / L_{\rm box})^3 / 3$, where $N_i$ is the number of voids with radius between $R_{i-1}$ and $R_i$ and $\langle R \rangle_i = (R_{i-1} + R_i) / 2$;
the variance $\Sigma_i$ is given by the Poissonian uncertainty of the measured multiplicity function, $\Sigma_i = N_i  \left[4 \pi (\langle R \rangle_i / L_{\rm box})^3 / 3 \right]^2$. The prior is a flat distribution over the 3 parameters explored. Note that the radius bins explored in this analysis are thin, with a thickness ranging from $0.5$ to 1 $h^{-1}$Mpc. 
We repeat the analysis exploring several minimum void radius values. For each minimum radius, we verify to be in the regime in which the approximated analytical multiplicity function is an accurate approximation of the exact one corresponding to the effective barrier. To do so, we compare the best-fit approximated multiplicity function, Eq.~\eqref{eq:multipl_integr_approx}, with the numerical one obtained with the same moving barrier, Eq.~\eqref{eq:delta_sigle_cholesky}--\eqref{eq:num_f}. According to the precision of the analytical approximation Eq.~\eqref{eq:multipl_integr_approx} and to the uncertainty of the VSF from ICs, we select a minimum radius in which the analytical approximation can be safely used, in a very conservative way, listed in the last column of Tab.~\ref{tab:Rmin}. Then we select the parameter space region to be further explored as the region in which the logarithm of the posterior distribution is larger that the maximum value minus 5: $\theta$ s.t. $\log(p_{\rm max}) - \log\big(p(x,\theta)\big) > 5$, where $\theta = (\alpha,\beta,\gamma)$ and $x$ the measured VSF. We then use this region in the parameter space as the flat prior for the analysis with the numerical multiplicity function, Eq.~\eqref{eq:delta_sigle_cholesky}--\eqref{eq:num_f}. We repeat this procedure for all the void thresholds considered.
It is worth noting that both Eq.~\eqref{eq:f_new_final} and Eq.~\eqref{eq:multipl_integr_approx} are able to fit the VSF from simulation ICs even at small radii, but beyond the radius in which the analytical approximation breaks, this is no more representative of the first crossing solution of a generic moving barrier. This means that even if in principle we can use Eqs.~\eqref{eq:multipl_integr_approx}, to fit the parameters of the moving barrier Eq.~\eqref{eq:moving_b}, at scales smaller than the radius at which the analytical approximation breaks, we lose the map between this function and the effective barrier. Since the aim of Sec.~\ref{sec:effective_barrier} is to find the effective moving barrier, we proceed to measure it using the numerical multiplicity function.

\begin{table}[t!]
\centering
\begin{tabular}{ccccccccccc}
\toprule
\midrule
$\delta(z_{\rm IC})$ &&  $R_{\rm min}(2000)$  && $N_{\rm v}(2000)$  &&  $R_{\rm min}(500)$ &&  $N_{\rm v}(500)$  &&  $R^{\rm A}_{\rm min}$  \\
\midrule
-0.035  &&  11  && 361 &&  2.5 && 92965 &&  4  \\
-0.023  &&  17  && 419 &&  5  && 14900 &&  7  \\
-0.016  &&  17  && 9586 &&  7  && 6703 &&  11  \\
-0.011  &&  19  && 18419 &&  9 && 3601 && 19  \\ 
-0.008  &&  22  && 16437 &&  11 && 2031 && 28  \\
-0.0064 &&  30  && 6556 &&  9  && 3198 && 35  \\
-0.005  &&  39  && 3185 &&  12 && 1386 && 45  \\ 
\midrule
\bottomrule
\end{tabular}
\caption{Minimum radii for each threshold explored. The first column, $\delta(z_{\rm IC})$, lists the actual density contrast in the simulation ICs at $z_{\rm IC}=99$; the second and fourth columns, $R_{\rm min}(2000)$ and $R_{\rm min}(500)$, list the minimum radii used for the analysis with the numerical multiplicity function, which contains to $\sim1.5-2.5$ times the number of particles associated to the Poissonian noise (see text); the third and fifth columns, $N_{\rm v}(2000)$ and $N_{\rm v}(500)$, list the number of voids in the simulation ICs with radius larger then the minimum one; the last column, $R^{\rm A}_{\rm min}$, lists the minimum radius used for the MCMC analysis with the analytical approximation, Eq.~\eqref{eq:multipl_integr_approx}, chosen in a conservative way. All radii are in $h^{-1}{\rm Mpc}$ unit.}
\label{tab:Rmin}
\end{table}

To perform the MCMC up to the smallest possible radius allowed by ICs resolution, we use the numerical solution of the first crossing problem, Eqs.~\eqref{eq:delta_sigle_cholesky}--\eqref{eq:num_f}, implemented with the 3-parameters barrier Eq.~\eqref{eq:moving_b}. To obtain a numerical multiplicity function with enough precision to be used in such MCMC, we use $10^8$ MC realisations of Eq.~\eqref{eq:delta_sigle_cholesky} for each $\alpha$, $\beta$, and $\gamma$ values, computed with a binning radius size at least 4 times smaller than the binning used for the VFS measured in simulations ICs. This guarantees a numerical precision that is, at worst, less than $1/10$ the Poissonian uncertainty of the measured VSF. 
Since this is computationally expensive for a MCMC analysis, we interpolate the numerical multiplicity function over a sample of $\alpha$, $\beta$, and $\gamma$ values. The computational cost scales linearly with the number of realisations performed and quadratically with the number of radius bins considered. For reference, with the above specifications, the first crossing distribution computed with $10^8$ realisations in a random position in the parameter space within domino for the $\delta_{\rm v}=-0.62$ case, and 109 radius bins, takes around 13 seconds on a core with 48 CPUs at 2.90GHz. 
To select the range in a convenient way, we produce a large number of numerical multiplicity over a random sapling of $\alpha$, $\beta$, and $\gamma$, selected within the parameter space region delimited by the MCMC analysis involving the analytical multiplicity function, as described before. 

At each MCMC step, we interpolate over this sample finding the closest tetrad in which the point of coordinates ($\alpha$, $\beta$, $\gamma$) is embedded (which in most of the cases does not correspond to the 4 closest sample points).  For interpolating, we weight the numerical multiplicity function corresponding to each point using the barycenter coordinates. The likelihood is the same as Eq.~\eqref{eq:likelihood}, however, the theoretical model is now obtained by numerically integrating over the 4 theoretical radius bins composing one data radius bin, i.e. $n^T_i(\Theta) = \sum_{j=1}^4 N_\times (R_j) / N_{\rm tot}$, with $N_{\rm tot}=10^8$, where $N_\times (R_j)$ is the number of crossing interpolated with the barycenter coordinates. In addition, the variance is now the square sum of the numerical uncertainty of the theoretical model and the Poissonian uncertainty of measurements from simulations: $\Sigma_i = N_i  \left[4 \pi (\langle R \rangle_i / L_{\rm box})^3 / 3 \right]^2 + \sum_{j=1}^4 N_\times (R_j) / N_{\rm tot}^2$. We verified that, for this specific application, this interpolation method is more accurate than the standard ones over a grid. Moreover, this approach allows to simply increase the sample resolution if necessary, by producing more realisations at random positions within the same parameter space region, without worrying about the grid.

As a last consideration, it is worth noting that the moving barrier shape, Eq.~\ref{eq:moving_b}, is robust in modelling the Lagrangian VSF measured from simulation ICs. In particular, the five larger radius bins are always enough to constrain the 3-parameter barrier and to predict the VSF down to the smaller radii. However, using all the available radius bins, the uncertainty in the posterior decreases. \\
All the MCMC analyses are performed using the {\tt emcee} sampler~\citep{emcee}.

\section{Redshift dependence of the HMF}\label{ap:redshift_evolution}

\begin{figure}[t!]
\centering
\includegraphics[width=0.67\textwidth]
{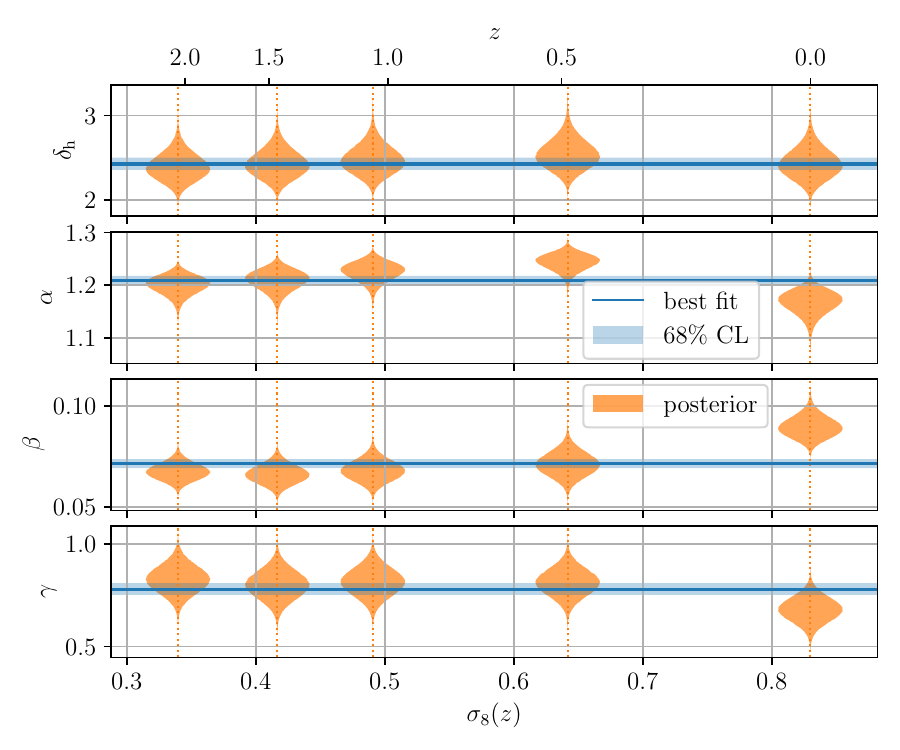}
\caption{Dependence of the halo formation threshold $\delta_{\rm h}$ and the moving barrier parameter parameters of Eq.~\eqref{eq:moving_b} as a function of $\sigma_8(z)$. The upper axis shows the corresponding redshift. Orange violins show the posterior distribution obtained at each DEMNUni snapshot considered, $z=0$, 0.47, 1.05, 1.46, 2.05. Blue solid lines and dashed areas show the best-fit and 68\% CL of the scaling relations described in the text.
}
\label{fig:params_wrt_sig8}
\end{figure}

In this Section we discuss the redshift dependence of the Eulerian HMF, as described in Sec.~\ref{sec:eulerian_mapping}, obtained using Eq.~\eqref{eq:moving_b_params_val} as prior to fit the formation barrier $\delta_{\rm h}$. The redshift dependence of the theoretical HMF is better expressed as a function of $\sigma_8(z) = \sqrt{C(R,R)}$ at a given redshift, with $R=8\,h^{-1}{\rm Mpc}$. In particular, $\sigma_8(z) = \sigma_8(z=0) D_0(z) / D_0(z=0)$, where $D_0(z)$ is the linear growth factor of perturbations. In this way, the normalisation effects of the Lagrangian space and any redshift dependence are automatically included. 
Nevertheless, Fig.~\ref{fig:params_wrt_sig8} shows that all the parameters of the effective barrier, Eq.~\eqref{eq:moving_b} and $\delta_{\rm h}$ can be consistently considered redshift independent. 
The orange violins shows the posterior of the parameter at the $\sigma_8(z)$ values corresponding to the redshift values explored, $z=0$, 0.47, 1.05, 1.46, 2.05. Blue lines show the best-fit value and the shaded area the 68\% CL.  
The best-fit values are the following:
\begin{align}\label{eq:HMF_params_s8}
\delta_{\rm h} &= 2.424 \pm^{0.076}_{0.071}  &&& \alpha &= 1.209 \pm^{0.008}_{0.007}  \\
\beta &= -0.072 \pm^{0.022}_{0.022} &&& \gamma &= 0.776 \pm^{0.033}_{0.025}\,. \nonumber
\end{align}
It is worth noting that redshift independent barrier parameters entail a self-similar HMF, i.e. an HMF that depends on $S=\langle \delta^2 \rangle$ (and $D$) values only.
Note that the Lagrangian space is the initial density field linearly evolved up to the redshift of interest with the linear growth factor. This means that the linear formation threshold $\delta_{\rm h}$ does not depend on the redshift, but the linear growth factor used to evolve the Lagrangian space does. In computations, this can be considered alternatively by using the linear power spectrum of the redshift considered, or using the power spectrum at a reference redshift, e.g. $z=0$, and extrapolating the formation threshold at the redshift of interest: $\delta_{\rm h} \rightarrow \delta_{\rm h} D_0(z=0) / D_0(z)$ ~\citepalias{sheth_mo_tormen_2001}. According to the same argument, the effective moving barrier can be used either with the linear power spectrum at the redshift considered, or with the power spectrum at a reference redshift and extrapolating the barrier, which means $\alpha \rightarrow \alpha D_0(z=0) / D_0(z)$, where we considered $z=0$ as the reference redshift. Moreover, any parameter multiplied or divided by $\sigma$, or more generally by any quantity derived from the power spectrum, must be corrected for the growth factor. Therefore, also $\beta \rightarrow \beta D_0(z=0) / D_0(z)$.

\begin{figure}[t!]
\centering
\includegraphics[width=0.95\textwidth]{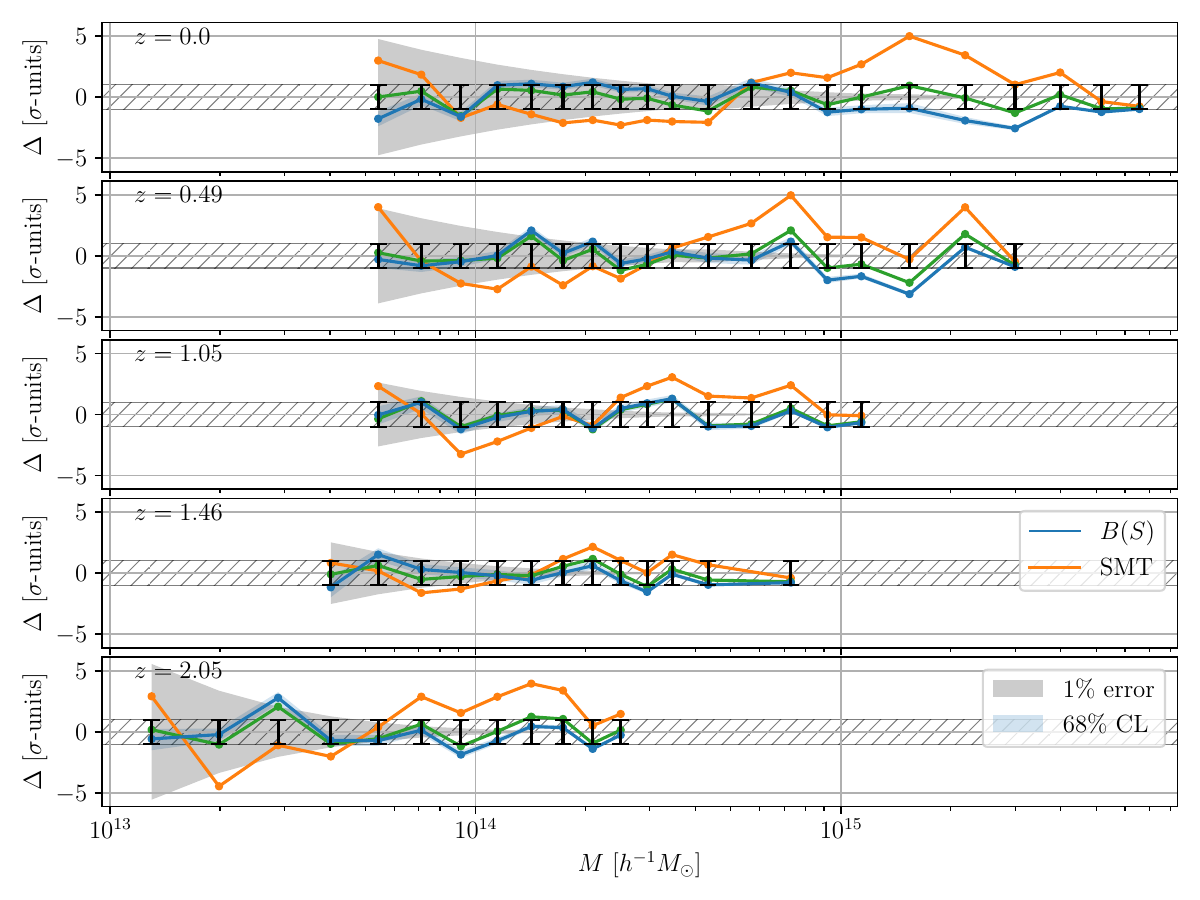}
\caption{Relative difference between theoretical
HMFs and with respect to the HMF measured in the DEMNUni in unit of Poissonian uncertainty:
Eulerian spherical mapping of the effective barrier approach implemented in this work (blue), Eqs.~\eqref{eq:multipl_integr_approx},~\eqref{eq:moving_b},  fitting the halo threshold from Eq.~\eqref{eq:moving_b_params_val}; \citetalias{sheth_mo_tormen_2001} best-fit model (orange); Tinker et al. 2008~\citep{tinker_2008} best-fit model (green).
Each panel shows a different redshift, $z=0$, 0.47, 1.05, 1.46, 2.05. Errorbars and hatched area show $\pm 1\sigma$ interval, dashed area shows the $\pm 1\%$ uncertainty.}
\label{fig:HMF_EU_all_z}
\end{figure}

The blue curve in Fig.~\ref{fig:HMF_EU_all_z} shows the relative difference between the Eulerian spherical mapping of the effective barrier approach implemented in this work (blue), Eqs.~\eqref{eq:multipl_integr_approx},~\eqref{eq:moving_b}, fitting the halo threshold from Eq.~\eqref{eq:moving_b_params_val} as described in Sec.~\ref{sec:eulerian_mapping}. 
Each panel shows a different redshift, corresponding to the DEMNUni snapshots considered. The error bars and the hatched area show $\pm 1\sigma$ interval, dashed area shows the $\pm 1\%$ uncertainty. The agreement is almost always within $1\sigma$. Similarly to the case of fitting the moving barrier for the Lagrangian VSF, Appendix~\ref{ap:mcmc_barrier}, the model presented here is robust in modelling the Eulerian HMF measured from simulation. As in the previous case, the posterior distributions of the considered parameter are stable from the larger scale explored down to the resolution limit of the simulation. 
For comparison, the best-fit of the \citetalias{sheth_mo_tormen_2001} model (orange lines) and the best-fit of the 4-parameters Tinker et al. 2008~\citep{tinker_2008} model are shown. It can be noticed that the agreement of our model with the HMF measured in simulation is almost always as good as the Tinker et al. 2008~\citep{tinker_2008} one. The main differences are the clearer physical interpretation and the smaller parameter space of the model presented in this work. Moreover, our model converges much faster in the MCMC fit than the 4-parameters Tinker et al. 2008~\citep{tinker_2008} model. Note that the multiplicity functions of Eq.~\eqref{eq:f_new_final} and Eq.~\eqref{eq:multipl_integr_approx} can be also used as a versatile fitting function with as many fitting parameters as desired, which modify $B_s$ and $B'_S$ quantities.

\end{document}